\documentclass[useAMS,usenatbib]{mn2e}
\usepackage{amsmath}
\usepackage{natbib}
\usepackage{graphicx}
\usepackage{color}
\usepackage{rotating}
\usepackage{nicefrac}
\usepackage{txfonts}
\usepackage{subfigure}
\usepackage[percent]{overpic}
\usepackage{xspace}
\usepackage{url}
\usepackage{chemarrow}
\usepackage{amssymb}
\usepackage[percent]{overpic}
\title{Embryo impacts and gas giant mergers II:
Diversity of Hot Jupiters' internal structure}
\author[Shang-Fei Liu, Craig B.~Agnor, D.~N.~C.~Lin and Shu-Lin Li]{Shang-Fei Liu$^{1,2}$, Craig B.~Agnor$^{3}$, D.~N.~C.~Lin$^{1,4,5}$ and Shu-Lin Li$^{1,6}$\\
\\
$^1$ Kavli Institute for Astronomy and Astrophysics and Department of Astronomy, 
Peking University, Beijing 100871, China; E-mail: \url{liushangfei@pku.edu.cn}\\
$^2$ Department of Earth and Planetary Sciences,
University of California, Santa Cruz, CA 95064, USA\\
$^3$ Astronomy Unit, School of Physics and Astronomy,  
Queen Mary University of London, United Kingdom\\
$^4$ Department of Astronomy and Astrophysics,
University of California, Santa Cruz, CA 95064, USA\\
$^5$ Institute for Advanced Studies, Tsinghua University,
Beijing 100084, China\\
$^6$ National Astronomical Observatory of China, Chinese Academy of Sciences
Beijing 100012, China}
 
\begin{document}
\maketitle
\begin{abstract}
We consider the origin of compact, short-period, Jupiter-mass planets.
We propose that their diverse structure is caused by giant impacts of embryos 
and super-Earths or mergers with other gas giants during the formation 
and evolution of these hot Jupiters. Through a series of numerical 
simulations, we show that typical head-on collisions generally lead 
to total coalescence of impinging gas giants. Although extremely 
energetic collisions can disintegrate the envelope of gas giants, 
these events seldom occur. During oblique
and moderately energetic collisions, the merger products retain higher
fraction of the colliders' cores than their envelopes. They can also
deposit considerable amount of spin angular momentum to the gas
giants and desynchronize their spins from their orbital mean motion.
We find that the oblateness of gas giants can be used to infer the impact history.
Subsequent dissipation of stellar tide inside the planets' envelope
can lead to runaway inflation and potentially a substantial loss of gas through Roche-lobe overflow. The impact of super-Earths
on parabolic orbits can also enlarge gas giant planets' envelope
and elevates their tidal dissipation rate over $\sim $ 100 Myr time
scale. Since giant impacts occur stochastically with a range of
impactor sizes and energies,
their diverse outcomes may account for the dispersion in the mass-radius relationship of hot Jupiters.

\end{abstract}


\section{Introduction}
In the conventional sequential accretion hypothesis (SAH), gas giant
planets acquire critical-mass ($\sim 10 M_\oplus$) cores prior to the
onset of efficient gas accretion \citep{Pollack:1996zr, Ida:2004ko}.
This scenario can account for many observational properties
of Jupiter, Saturn, and extra-solar planets \citep{Schlaufman:2009bh}.
However, the observed large dispersion in the internal structures of gas giant planets was not anticipated, even taking into account the fact that the assumptions they made to get this critical value may not be universal.

In the first paper of this series \citep[][hereafter Paper I]{Li:2010fk}, 
we attribute the dichotomy between the mass of Jupiter and Saturn's 
cores to giant impacts by embryos and mergers of gas giants during 
their formation and early dynamical evolution. Using two
numerical methods, we simulated a series of models with parabolic
collisions and show that these collisions generally lead to the
accretion of embryos and coalescence of gas giants. We also show that
massive super-Earth impactors can survive passage deep into the envelope of a gas giant
and release energy near the core.  This may drive intense convective
motion in the planet and core erosion. In contrast, sub-Earths embryos
disintegrate higher in the gaseous envelope and their fragments will dissolve in the envelope and contribute to a heavy element enrichment.

\begin{figure}
  \centering
  \includegraphics[width= \linewidth,clip=true]{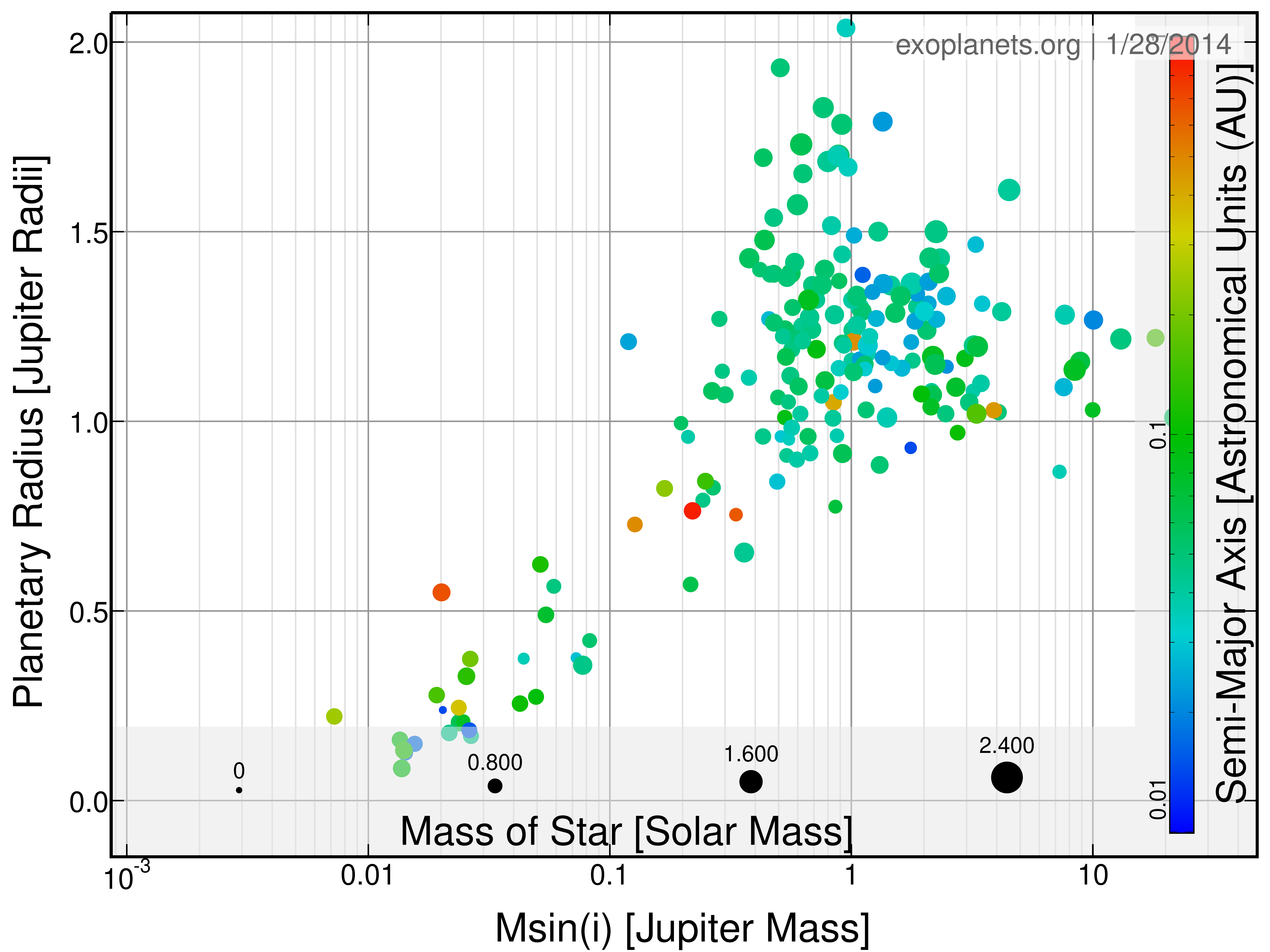}
\caption{The mass-radius distribution of transiting exoplanets color coded 
with their semi-major axes. The size of symbols indicate masses of their host 
stars. Data were taken from the Exoplanet Data Explorer 
(http://exoplanets.org/) on Jan. 28, 2014.}
\label{fig:f0}
\end{figure}

Transit surveys suggest that there is a remarkable diversity in hot Jupiters' radii. Extrasolar planets
with masses comparable to Jupiter range in size by a factor of more
than two (Figure \ref{fig:f0}).  Some of
these planets ({\it e.g.} HD\,209458 b) have Jupiter mass ($M_{\rm J}$) but
more than 30 \% larger radius. Several planets (such as Corot 8 b and
WASP 29 b) have masses comparable to Saturn, but radii $R_{\rm p}$ that are
much smaller than Saturn's.  These compact planets may require the 
presence of massive cores ({\it e.g.} up to 70-$M_\oplus$ core is
needed for HD\,149026 b). The main focus of this paper is to apply
the giant impact and merger scenario (GIMs) to account for large
dispersion in hot Jupiters' radii.

In order to construct this scenario, we briefly cite observational
evidence for the structural and atmospheric diversity of hot Jupiters. 
Through a brief recapitulation of their origin, we present arguments 
in \S2, to suggest that GIMs occur frequently during 
and after orbital migration to the proximity of their host 
stars. In \S3, we used two complementary approaches:  a Lagrangian 
smoothed particle hydrodynamics (SPH) 
code, and the Eulerian adaptive mesh hydrodynamics code FLASH 
\citep{Fryxell:2000kx} to simulate both head-on and oblique 
collisions between a Saturn-mass giant planet and a 10-$M_\oplus$ 
or 25-$M_\oplus$ embryo. These models extends the previous models 
to the high energy limit. We show that most hyperbolic oblique 
encounters deposit spin angular momentum to gas giants. Highly 
energetic collisions can directly lead to the substantial loss of 
envelope gas beyond their Roche radii. SPH and FLASH simulations 
of merger between two Saturn-like gas giants are also presented. 

In order to expand the dynamical range for density and investigate the long term evolution
in our simulated models, we adopted, in paper I, a one-dimensional Lagrangian hydrodynamic
(LHD) scheme which can be used to study the effect of tidal
interaction of inflated merger products with the host stars. With
LHD, we show in \S4, the energy released from modest giant impacts can
also induce the expansion of the gas giants' envelope. Subsequent
dissipation of stellar tidal perturbation in the gas giants may lead
to runaway inflation.  Since gas in the envelope is preferentially 
lost, these encounters lead to the enhancement of the planet's 
metallicity and asymptotic compact size.  Finally, we summarize our 
results and discuss their observational implications in \S5.

\section{Giant impact history of close-in gas giants.}
\label{sec:justify}

In paper I, we discuss the possibility of GIMs for long-period gas
giants in the proximity of the birth place as an avenue for the
enrichment of their envelope and a cause for structural diversity
between Jupiter and Saturn. It is difficult to directly detect the
metallicity and accurately determine the internal structure of
extrasolar planets.  However, if the metallicity ([Fe/H]), envelope and core masses
($M_{\rm env}$ and $M_{\rm core}$), the orbit and spin state of any planet are known, it is possible to construct
its contraction sequence, taking into account  stellar irradiation and
tidal dissipation \citep{Bodenheimer:2001qf, Bodenheimer:2003gd}. Although the
  interior structure of extrasolar gas giant planets remains uncertain
  and is an area of active research, in this work we simply ignore
  complicating factors, assume a core-envelope structure and assume that the
  core of any planet can be inferred from its measured radius ($R_{\rm
    P}$) and the age of its host star ($\tau_\ast$).

As of this writing there are more than 200 transiting gas giants with
measured masses
$M_{\rm p}$ and radii
$R_{\rm p}$ \footnote{see, e.g., 
\url{http://exoplanets.org/}.}.
Due to an observational selection effect, most of the known transiting
gas giant planets are close-in planets known as hot Jupiters
with periods and masses in the range of a few days and Jupiter mass
respectively. The diversity in the value of their measured radii $R_{\rm P}$
and densities motivated us to invoke the GIM scenario.

\subsection{Possible causes for inflated hot Jupiters.} 
\label{sec:hotjup}
Most of the transiting hot Jupiters have $R_{\rm p}$'s consistent with those
expected from models constructed for their $M_{\rm p}$ including cores with 
$M_{\rm core} \sim 10-20~M_\oplus$. However, there are several transiting
planets, such as HD\,209458b, Tres-4 \citep{Mandushev:2007qa}, and 
WASP-12b \citep{Li:2010lr} which
have observationally inferred radii much larger than that of Jupiter
($R_{\rm J}$) even though their mass is comparable to that of Jupiter. For
the special class of inflated hot Jupiters, suggestions for their
unusually low densities include:

\noindent
1) the thermal contraction of these planets is suppressed by the
stellar irradiation \citep{Burrows:2000ye, Burrows:2004tw}, 

\noindent
2) they are inflated by additional heating sources such as the
dissipation due to the ongoing stellar tidal perturbation
\citep{Bodenheimer:2001qf},

\noindent
3) irradiation-driven circulation \citep{Showman:2002ff}, 
or associated Ohmic dissipation \citep{Batygin:2010ys,Wu:2013kq}

\noindent
4) the sedimentation of heavy elements \citep{Baraffe:2008fu}. 

Most of these propositions have some difficulties in accounting for
the difference between the exceptions and the other ``normal'' gas
giant planets. Here we suggest another scenario.
Following  Paper I, we suggest 
that the energy dissipation during GIM's and the tidal dissipation 
thereafter may be adequate to inflate some gas giant planets.
The stochastic nature of GIM's provides a possible explanation for the
wide dispersion $M_{\rm p}-R_{\rm p}$ distribution of hot Jupiters.

\subsection{Hot Jupiters with massive cores or most metallic compositions.}
\label{sub:hotjupearth}

In paper I we showed that GIM's may lead to transitory
planetary inflation.  Merger events of two or more gas giants and
multiple impacts of earth-mass embryos may also lead to the formation
of massive cores and the heavy metallicity contamination of planetary
envelopes. In this paper, we explore the possibility that giant impacts and
mergers may produce mature, compact, hot Jupiters. 

HD\,149026 b has an observed radius of $R_{\rm p} =0.73 R_{\rm J}$ and the 
mass of Saturn \citep{Sato:2005uq}. 
Structural analysis indicates that more than half ($\sim 70~M_\oplus$)
of its total mass may be contained within a core of heavy elements
\citep{Sato:2005uq}. Later models suggest that it is also possible for
heavy elements to distribute throughout the interior of the planet
\citep{Ikoma:2006kx, Baraffe:2008fu}. The extraordinary compact radius of
HD\,149026 b poses a challenge to both gravitational instability and
sequential accretion hypothesis (SAH) for gas giant planet formation.

The required metallicity of HD\,149026 b is more than an order of magnitude
larger than that of its star (which has a mass $M_\ast =1.3~M_\odot$ and
metallicity [Fe/H] = 0.36). If this planet formed through gravitational
instability \citep{Boss:1997bh}, this heavy elemental concentration would 
require either an enormous loss of its original gaseous envelope or an 
extensive acquisition of heavy elemental material through the post-formation 
accretion of solid planetesimals and proto-planetary embryos.

If this planet acquired its present-day structure through core accretion
\citep{Pollack:1996zr} such a massive core would be
attainable in a gas-rich environment beyond the snow-line provided it can avoid initial dynamical
isolation. In principle, at several AU's from its host star, the isolation mass
$M_{\rm iso}$ of proto-planetary solid embryos can reach this massive level
provided the local surface density of heavy elements $\Sigma_{\rm g}$ is more than an
order of magnitude larger than that inferred from the MMN model. In the disk
around a host star with solar metallicity, the inferred $\Sigma_{\rm g}$ would render
the augmented disk to be gravitationally unstable.

However, if the core-building material is able to congregate in some
confined locations such as the snow line, solid embryos with very
large $M_{\rm iso}$ would be attainable in a gravitational stable
disk, prior to the accretion of a relatively massive envelope. An
additional requirement is that the planetesimal accretion rate must
exceed the gas accretion rate until the core has already acquired most
of its present-day mass. With its massive core, this
requirement may be difficult to accomplish for HD\,149026 b because the
heat loss is efficient and the gas accretion rate are likely to be
large around any cores with $M_{\rm core} > 20 M_\oplus$. 

During the phase of gas accretion, the protoplanets' feeding zones
expand with their masses. Early models for core accretion are
constructed under the assumption that all embryos in the expanding
feeding zone are accreted by the growing protoplanet
\citep{Pollack:1996zr}. However, these embryos have a tendency to migrate
away from the protoplanets and form gaps around them during their initial
phase of modest gas accretion. Only when the gas giants
rapidly accrete massive envelopes do the orbits of their nearby residual
embryos become destabilized and a modest fraction of the
orbit-crossing embryos would merge with them \citep{Zhou:2007vn, 
Shiraishi:2008ys}.  However, the total mass  
of HD\,149026 b's gaseous envelope is smaller than that contain in the heavy elements.

The above discussion indicates that the formation of some isolated gas
giants with massive cores cannot be ruled out {\it a priori}. But
these compact planets are apparently exceptions rather than the general rule.
Around a host star OGLE-TR-132, a transiting planet has been found
with similar $M_{\rm P}$ and $a$ as those of HD\,149026 b. Despite the nearly
identical [Fe/H] for both host stars, OGLE-TR-132 b has a radius nearly
twice that of HD\,149026 b \citep{Gillon:2007fk}. Presumably, OGLE-TR-132 b
has a much smaller core mass than HD\,149026 b. The large discrepancy
between these two same-mass planets around similar host stars again
points to stochastic origins of their internal structures. The large
dispersion in the average density of known gas giants (Figure 1) is 
further evidence of this structure diversity. 

Here we adopt the proposition and examine the possibility that the
massive core of compact planets such as HD\,149026 b are the byproduct of 
one or more giant impacts and mergers \citep{Sato:2005uq, 
Ikoma:2006kx} in close proximity to its host star. Whereas those 
planets with unusually large
radii may have been struck recently (within the last $\sim 100$ Myr)
by modest-mass embryos (with masses up to $\sim M_\oplus$), the
impactors for this compact planet are either another gas giant or
very massive (up to $\sim 10-20 M_\oplus$) embryos. We also suggest
that these collisions occur so long ($>$ 1 Gyr) ago that the internal
structure of this planet has had adequate time to readjust to a new
thermal equilibrium.

\subsection{Availabilities of residual embryos as 
potential impactors of emerging gas giants}
\label{sec:origin}
We first consider the possibility of giant impacts during the 
formation of gas giants.  In typical disks around classical 
T Tauri stars, super-Earth embryos are not sufficiently 
massive to open gaps \citep{Lin:1993hl}.  Nonetheless, 
they tidally interact with the disk and undergo type I 
migration\citep{Goldreich:1980vn, Ward:1984ys, Ward:1997mw}.  
Their migration time scale and direction are determined 
by the sum of the planets' corotation and Lindblad torques 
on the disk and by the disk structure
\citep{Paardekooper:2011rr}.  There is a tendency for the super-Earth
embryos to converge toward some trapping radii  
where their enhenced surface density promotes the embryos' 
oligarchic growth and the emergence of cores with sufficient
mass to initiate efficient gas accretion.  As they rapidly 
gain mass, emerging protoplanets destabilize the orbits of 
neighboring residual sub-critical embryos \citep{Zhou:2007vn}.  Some
of these embryos collide with the protoplanets (see paper I).  

With sufficient masses, proto gas giants open gaps. In extended 
protostellar disks, gas giants form interior to the half mass 
radius of the disk and undergo inward type II migration and become 
hot Jupiters\citep{Lin:1996ey}. Along the way, the migrating
gas giants capture residual embryos in their sweeping mean 
motion resonances \citep{Zhou:2005pr}. 

Although the relatively low-mass embryos have not migrated 
extensively on their own, they are being shepherded to spiral 
toward their stars with the migrating gas giants (cf
\citealt{Yu:2001sh}). This snow-plough effect has led to 
the formation of the resonant gas giant planets systems around 
55 Cnc and GJ 876 \citep{Lee:2002ef}. This process can excite 
embryos' eccentricity, lead to orbit crossing, and enhance 
the possibility of giant impacts \citep{Zhou:2005pr, 
Fogg:2007nx, Mandell:2007kl}.

Many gas giants are members of multiple planet systems.  During the 
stage when two or more gas giants capture each other into their mutual
mean motion resonances (such as those around GJ 876 and 55 Cnc) or enter into secular
resonances (such as those around Ups And), their apsidal precession
frequencies change greatly \citep{Murray:1999le}. The location of their
secular resonances also sweeps over extensively wide regions during
the depletion of the solar nebula \citep{Ward:1981wb}, or during later
gas-free planetesimal-driven or instability-driven giant planet migration \citep{Agnor:2012dn}. These
resonances may excite eccentricities of super-Earths and result in
giant impacts with gas giants.

\citet{Ketchum:2011aa} studied an alternative scenario, in which super-Earths form and migrate inward after hot Jupiters complete inward migration and reside inside the inner edge of the disk. They found those super-Earths have a great chance to collide with the inner gas giant if the eccentricity damping is not efficient.

\subsection{Supply of potential impactors to hot Jupiters during 
the epoch of disk depletion.} 
\label{sec:superearths}

The trapping radius where super Earths converge depends the 
surface density and temperature distribution in the disk.
These quantities are also functions of disks' accretion rate,
effective viscosity, and stellar luminosity \citep{Garaud:2007oq}. 
In models with a standard $\alpha$ prescription, 
super Earths' trapping radius is located at the transition 
between the inner disk which is heated by viscous dissipation
and the outer disk where stellar irradiation provides the 
dominant thermal energy \citep{Kretke:2012ai}.  During the 
advanced stage of disk evolution when accretion diminishes
with gas depletion, the trapping radius contracts and
the magnetosphere truncation radius expands. Through this process,
super Earths may accumulate in the proximity of their host
stars.  

On the observational front, radial velocity and transit surveys 
indicate that solar-type stars bear super-Earths, with mass 
and period up to $\sim$ 20 $M_\oplus$ and a few months, are more 
common than those with Jupiter-mass gas giants.  Radial velocity 
survey with HARPS reports that it may be in the range 39-58\%,
though only a few of these claims have actually been published 
(Mayor et al 2009, Udry private communication). Although the 
eta-of-Earth survey \citep{Howard:2010bs} suggests a lower 
fraction of stars with super-Earths, it nonetheless indicates 
that the frequency of stars with short-period planets is 
a rapidly decreasing function of planet mass. The common 
existence of super-Earths is also suggested by many detections 
in the Kepler transit survey.  In many cases, these super-Earths 
are members of multiple-planet systems in which migration and 
dynamical interaction most likely have influenced their 
formation and evolution \citep{Papaloizou:2010qf}. 

After the disk depletion, mutual secular perturbation between 
co-existing hot Jupiters and close-in super Earths leads to 
eccentricity excitation and eventually dynamical instability.
Numerical simulations \citep{Zhou:2007fk} indicate that multiple 
planets, with equal masses $M_{\rm p}$ and a normalized separation $k_0 
= \Delta a / R_{\rm H}$ where $a$, $\Delta a$, and $R_{\rm H} = a(2 M_{\rm p} / 
3 M_\ast)^{1/3}$ are the semi major axis, separation, and Hill 
radius, undergo orbit crossing on a time scale 
\begin{equation}
{\rm log} (T_{\rm c}/P_{\rm k}) \simeq -5 + 2.2 k_0 \,,
\label{eq:tcross}
\end{equation}
where $P_{\rm k}$ is the mean orbital period.  

In \S2 Paper I, we have indicated that self
gravity of protostellar disks introduces precession in the eccentric
orbits of gas giants. Secular resonances occur in regions of disks
where precession rates due to the gravitational perturbation of these
planets matches with their precession rate due to the disk
potential. As $\Sigma_{\rm g}$ decreases, these secular resonances sweep
across them. Passage of the secular resonances excites the
eccentricity of residual planets and super-Earths \citep{Ward:1981wb,
Nagasawa:2005oa} including those with short periods (such as
$\upsilon$ And b and $\mu$ Ari b \citep{Nagasawa:2005xz,
Zhou:2005pr}). This process can also lead to orbit crossings which
destabilize multiple planet systems and induce potential giant impact
events\citep{Thommes:2008jb}.

\subsection{Hot Jupiters' potential impactors around mature stars} 

In compact systems of multiple hot Jupiters and close-in super-Earths
the crossing time scale may be lengthened by the stabilizing contributions
of relativistic precession and internal tidal dissipation 
\citep{Mardling:2004mq, Mardling:2007wm, Terquem:2007lp}. However, these
planets undergo further decay\citep{Novak:2003ta, Lee:2013ph} 
which may eventually lead to dynamical instabilities, 
orbit crossing, and eventually cohesive collisions \citep{Lin:1997ig}.

Compact systems of long-period planets also undergo orbit crossing, 
on the time scale of $T_{\rm c}$ and excite each other eccentricities
\citep{Weidenschilling:1996ij, Rasio:1996kx, Lin:1997ig, 
Zhou:2007fk, Juric:2008uq, Chatterjee:2008gd}.  Known gas giants 
with periods longer than 1-2 weeks have nearly uniform eccentricity 
distribution ranging from zero to nearly unity. Subsequent close 
encounters, secular chaos, Kozai resonances, and external 
perturbations \citep{Wu:2003lo, Wu:2007bs}, 
\citep{Laughlin:1998fv, Adams:2001dz, Spurzem:2009fu,
Nagasawa:2011dp, Wu:2011df}
may drive distant planets to highly eccentric orbits where they may
begin nearly parabolic close encounters with their host stars. It has been suggested
that a significant fraction of hot Jupiters, especially those with 
high obliquities, may be scattered to the proximity of their host 
stars with follow-up tidal circularization of their orbits 
\citep{Faber:2005sp, Fabrycky:2007kl, Narita:2007qa,
Winn:2007mi, Wolf:2007pi}. There is also a suggestion that the mass-period
distribution for the injected planets \citep{Nagasawa:2008ff} is consistent 
with that observed for hot Jupiters \citep{Marchi:2009lh}.
However, a fractional loss of envelope during their periastron passage 
may lead to the ejection of gas giants on parabolic orbits 
\citep{Guillochon:2011rt, Liu:2013uq}.  
Energy dissipation within the planets' envelope 
during the orbital circularization may also lead to substantial 
envelope inflation, mass loss, or total disruption \citep{Gu:2003ez}. 

Around stars with pre-existing short-period gas giants (which may have
migrated to the stellar proximity through disk-planet tidal
interaction), a fresh injection of both gas giants and embryos
raises the possibility of highly energetic merger and collision
events. In the following sections, we present models based on the
simulations of such events.

\section{SPH and grid-based simulations of Giant Impact and Merger 
of Gas Giants}
\label{sec:sph}

\subsection{Collisional Modelling Methods}
In paper I, we considered only head-on parabolic collisions between
embryos and gas giants. For long-period gas giants and for protoplanets
near their ice-line natal site, their local Keplerian velocity ($v_{\rm 
kep} \sim 10$ km s$^{-1}$) is generally smaller than their surface 
escape speed $v_{\rm esc} = (2 {\rm G} M_{\rm p} /R_{\rm p})^{1/2}$, so that their 
collisions are unlikely to be much more energetic than parabolic impacts. 
For hot Jupiters, however, the relative speed between (some parabolic) 
projectiles and target planets may exceed the local $v_{\rm kep}$ which
is $\geq$ 200 km s$^{-1}$. In anticipation on these possibilities,
we expand the range of impact speed in the simulated models presented
here. 

We model head-on and oblique collisions between planets using two complementary methods,
smoothed particle hydrodynamics (SPH) and an Eulerian grid-based code
FLASH. Our SPH code is a descendant of the one used in
\citet{Benz:1986mo} and includes self-gravity but neglects internal
strength. We use an ideal gas equation of state for the gaseous
envelope and the Tillotson equation of state for iron and basalt for
condensed materials. The Tillotson equation of state relates pressure, density and
 internal energy of condensed materials and models them through three regimes.
At low pressure this EOS captures the linear-shock particle relations
and at high pressure the EOS extrapolates to the Thomas-Fermi limit.
The Tillotson EOS also treats condensed material at low density as
either a high energy gas or low energy and pressure fluid.  Tillotson
 EOS parameters for particular materials (e.g., basalt, iron) are
 tuned to match results of laboratory experiments and theoretical limits \citep[see
Appendix A1 of ][ for a brief discussion]{Melosh:1989book}.

In addition, we use the Eulerian hydrodynamics code FLASH 
to construct several models with similar parameter used in SPH
simulations. FLASH is a parallel, multidimensional hydrodynamics 
code based on block-structured adaptive mesh refinement (AMR). The center of the computational domain is fixed to the center of mass of the system and is $4\times 10^{13}$ cm on a side. We have made a careful choice of the refinement criteria, making our smallest cells only $3.818\times 10^{7}$ cm in width, which is less than 7 parts in 1000 of the planet's radius. Combing the vast simulation domain with high resolution at the center, we are able to keep track of the envelope expansion after the impact and resolve the planet's dense core simultaneously.
Poisson equation is solved using a multipole expansion of the fluid 
with a maximum level at 40. Apart from from hydrodynamics and gravity, 
the two codes also differ in the equation of state (EOS). In FLASH 
simulations, we model the internal structure of the gas giant planet 
with composite polytropes with polytropic indices $n_1=0.5$ and $n_2=1$, which can characterize the distinct chemical composition of the core and envelope of a gas giant planet \citep{Liu:2013uq}. The embryo impactors are modeled with an $n=0.01$ polytrope with a central density $\rho_{\rm c} = 10$ g cm$^{-3}$. In this treatment, the polytropic constant $K$ is a free parameter, and the polytropic model can be determined by given values of $n$, $M$ and $R$ \citep{Kippenhahn:2013aa}. Here the choice of polytropic index $n=0.01$ over a more common value $n\simeq 1/3$ is made to give extra weight to the incompressible nature of the impactor. The polytropic index $n$ of the impactor has little effect on the post-impact thermal evolution of the target in the sense that the kinetic energy of the impactor dominates its internal energy and the target has a much larger heat capacity than that of the impactor. 

Our use of an ideal gas or polytrope EOS for the gaseous envelope
simplifies the computation at the expense of omitting the processing
of energy into internal degrees of freedom.  Shock heating of the
envelope resulting in dissociation and ionization of gas may act as a
sink to mechanical impact energy and affect the final
outcome of a collision \citep{Zeldovich:1967aa}.  As a result, our use of
simplified equations of state for the gaseous envelope may
increase the fraction of energy retained in mechanical kinetic energy
and the amount of material escaping a collision. Future work may
address the significance of this issue with a more sophisticated equation of
state and assess how of how this effect alters collisional processing
of giant planets. However, our SPH and FLASH collisions models serve to illustrate the
basic outcome morphologies and their implications for the collisional evolution of hot Jupiters.

This section is organized as follows. We first lay out our model of giant impact and merger. We then show that both codes produce consistent results, although some discrepancies are found. Finally, we will discussion the causes for these discrepancies and use our results to predict observational signatures.

\subsection{Range of impact speeds}
\label{sec:impactspeed}

In the limit that the orbits of two planets with $\Delta a \ll a$ 
marginally cross within each other's Roche radius, their relative speed is
\begin{equation}
\Delta v \sim v_{\rm kep} (a + \Delta a) ( 1 + {\Delta a / a}) - 
v_{\rm kep} (a) \simeq v_{\rm kep} (a) (\Delta a / 2 a) \,.
\label{eq:Deltav}
\end{equation}
Since gas giants' gravity also accelerates impactors, the impact speed
associated with GIM's is generally larger than this value. 

We first consider a system of planets which formed with nearly circular 
orbits.  After the gas in the natal disks is depleted, their $\Delta v$ 
grow and their orbits eventually cross each other on a time scale 
$T_{\rm c}$ (see Eq \ref{eq:tcross}) with $\Delta v \sim (G M_{\rm p} 
/ r_{\rm H})^{1/2} k_0/{\sqrt 12} \sim k_0/ 12^{1/3} ( M_{\rm p}/M_\ast)^{1/3}
v_{\rm kep}$. For systems of gas giants, formed at a few AU's with 
$k_0 \sim 6$, $T_{\rm c}$ is a few Gyr (ie comparable to the main sequence
life span of the solar type stars). For impactors with much lower masses, 
$\Delta a \sim 0.25 a$ and $\Delta v \sim 1-2$ km/s.  Their values may 
increase by 25\% for two gas giants with comparable masses.

Orbit crossing introduces finite collision probability rather than
certain GIM's. The collision time scale depends on the differential
orbital speed $\Delta v$ at which impactors enter into the gas giants'
$r_{\rm H}$.  For initial $k_0 > 1$, $\Delta v \sim (G M_{\rm p} / r_{\rm
H})^{1/2} \sim 3^{1/6} ( M_{\rm p}/M_\ast)^{1/3} v_{\rm kep} $ and  
most encounters do not necessarily lead to physical collisions. 
Nevertheless, since $\Delta v$ is much smaller than $v_{\rm esc}$, 
planet's mutual gravitational attraction would accelerate them to 
attain impact speeds for parabolic encounters {\it i.e.} 
$v_{\rm imp} \simeq v_{\rm esc}$.

Numerical experiments \citep{Zhou:2007fk} indicate that dynamical
instability can be suppressed by eccentricity damping.  In the 
proximity of their host stars, the eccentricity damping time scale
due to planets' internal tidal dissipation \citep{Goldreich:1966rp} is
\begin{equation}
\tau_e=\frac{4Q'_{\rm p}}{63 n_{\rm p}}\left(\frac{M_{\rm p}}{M_*}\right)
\left(\frac{a}{R_{\rm p}}\right)^5.
\label{eq:taue}
\end{equation}
The planets' quality factor $Q'_{\rm p} \sim 10^6$ and $\tau_e \sim 0.1$ Gyr  
are infered for most transiting hot Jupiters from their observed eccentricity-
mean motion ($e-n_p$) distribution \citep{Sasselov:2003zp, Ogilvie:2004rw}.  
Most close-in super-Earths have smaller $Q'_{\rm p}$ and $n_p$, and 
comparable $\tau_e$. From equations \ref{eq:tcross} and \ref{eq:taue}, we 
estimate $T_{\rm c} \sim \tau_e$ with $\Delta a \sim 0.3-0.4 a$ for 
$k_0 \sim 7$ for these planets. Their corresponding $\Delta v$ is 
comparable to their $v_{\rm esc}$ such that the speed ($v_{\rm imp}$) 
of their giant impacts may range from a fraction to several $v_{\rm esc}$.
For our numerical models, we consider this entire range of possibilities.

\subsection{Impact Orientation and Dynamics}

We describe collision dynamics using $v_{\rm imp}$ and impact angle
($\xi$). The magnitude of $v_{\rm imp}$ is a function of both the
relative velocity at infinity ($v_{\infty}$) and the two-body escape
velocity ($v_{\rm esc,2}$) with $v_{\rm imp}^2 = v_{\rm esc,2}^2 +
v_{\infty}^2$. For an encounter involving a target and impactor with
masses ($M_{\rm T}$, $M_{\rm I}$) and radii ($R_{\rm T}$, 
$R_{\rm I}$) respectively, the two-body escape velocity is
\begin{equation}
v_{\rm esc,2}^2 = \frac{2G (M_{\rm I} + M_{\rm T}) }
{(R_{\rm I} + R_{\rm T})} \,.
\end{equation}
In the limit $M_{\rm I} < < M_{\rm T}$ and $R_{\rm I} < < R_{\rm T}$,
$v_{\rm esc,2} \sim v_{\rm esc}$.  

We have performed simulations of a range of impact speeds with $v_{\rm
imp}/v_{\rm esc,2} = 1, 1.4, 3,$ and $5$, which is consistent with the
probable range of encounter velocities produced by multiple scattering events.

We use an impact angle ($\xi$) defined as the angle between the
relative position and velocity vectors when the surfaces of the two
bodies are in contact (i.e.~$\xi=0^{\circ}$ for a head-on collision and
$\xi =90^{\circ}$ for a grazing, tangential encounter). Assuming an
isotropic flux of impactors, the probability of a collision with $\xi$
in the range $\xi\rightarrow \xi+d\xi$ \citep{Shoemaker:1962wa} is
\begin{equation}
dP = 2\sin\xi\cos\xi\; d\xi \,.
\end{equation}
The angle $\xi=45^{\circ}$ is the median value
and the impact angles $\xi = 30^{\circ}$, $45^{\circ}$, $60^{\circ}$, and
$90^{\circ}$ divide the impact angle distribution into quartiles. We
have performed simulations of both head-on $\xi=0^{\circ}$ collisions
as well as more oblique collisions with impact angles of
$\xi=21^{\circ}$, $30^{\circ}$, $45^{\circ}$, and $60^{\circ}$.
SPH simulations were run for about $24$-hours of model time before
analyzing the result.  The simulations are first analyzed to determine
the particles that are contiguous with the initial gas giant's core.
Following that we determine which particles are gravitationally bound
in a two-body sense.  For FLASH simulations, we stop the simulations at about $2 \times 10^{5}$ s. The determination of contiguous and bound material is more challenging in a grid code, and we describe our approach to measure these quantities in Section \ref{sec:discrepancy}. Model
parameters of SPH and FLASH simulations and their results are summarized in Tables
\ref{tab:sph25}, \ref{tab:sph10}, and \ref{tab:saturn-100}.

\begin{figure}
  \centering
  \includegraphics[width= 0.90\linewidth,clip=true]{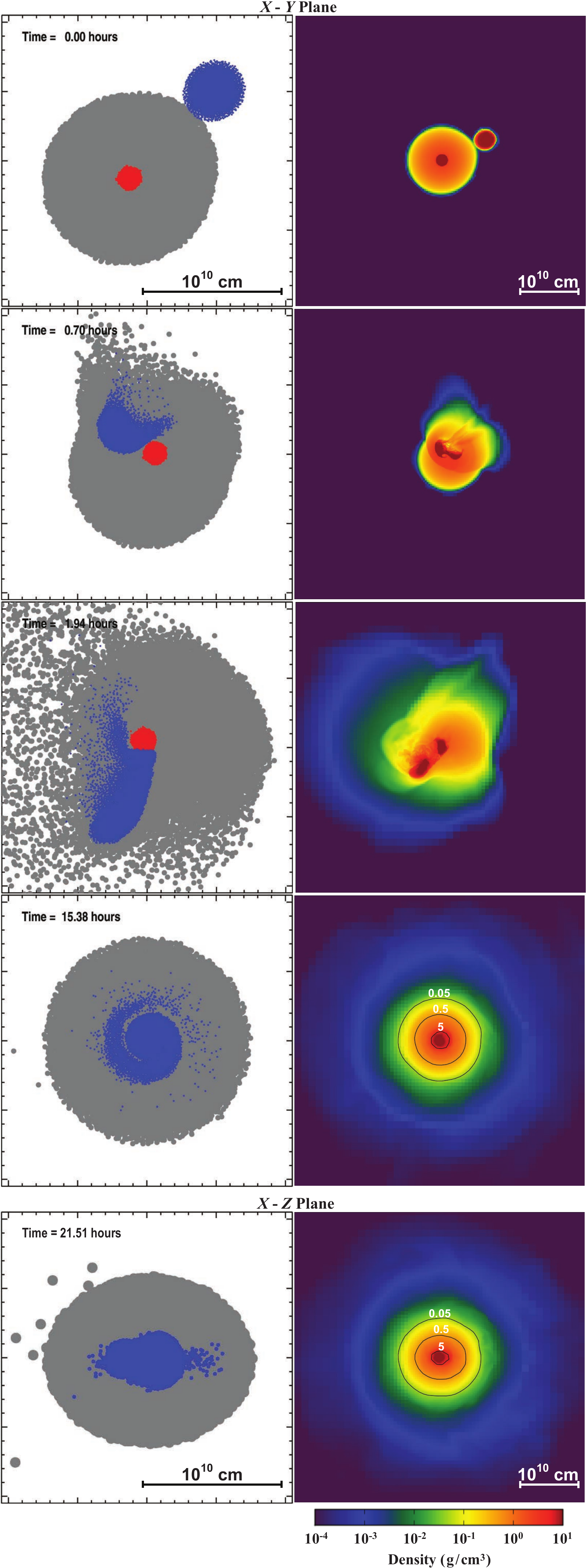}
\caption{Comparison of collision between a Saturn-like gas giant and 
a 25-$M_\oplus$ impactor. In the SPH model SA1c (left column), gas 
particles are shown as grey, and core particles (iron) are over plotted 
in red. The basalt impactor is over plotted in blue. In the FLASH 
model FA1c (right column), color denotes gas density, and the black 
lines in the last two frames are the contour lines at $\rho=5$, $0.5$ 
and $0.05$ ${\rm g \; cm}^{-3}$, respectively. The upper four rows 
are slices through the orbital plane (the initial motion is along the {\it X}-axis) and the lower one is the slices 
through the plane which perpendicular to the orbital plane.}
\label{fig:f1}
\end{figure}

\subsection{Collisions between a 25-$M_\oplus$ embryo
and a Saturn-mass gas giant} 

We simulate both head-on and oblique collisions
between a Saturn-like 100-$M_{\oplus}$ gas giant planet and a 25-$M_\oplus$ embryo.   
SPH model parameters and results of this series of simulations are
summarized in Table \ref{tab:sph25}. These results are the
generalization of the head-on parabolic collision in Paper I. In addition, four FLASH simulations with similar parameters are performed for comparison. Part of the analysis and visualization of the FLASH results in this work relies on the YT Python package \citep{Turk:2011aa}. 

\begin{table*}
\centering
\begin{tabular}{cccccccccc}
   \hline\hline Model & $\xi$ & $v_{\rm imp}/v_{\rm esc,2}$  &  $M_{\rm bm}$  & $M_{\rm bm,c}$  & $R_{\rm bm}/R_{\rm T}$  & $M_{\rm cc}$ & $M_{\rm cc,c}$  & $R_{\rm cc}/R_{\rm T}$ & $J_{\rm cc}/J_{\rm *}$ \\
    \hline
    SPH \\
  SA1a &  0 & 1.0 & 125.0 &  35.0 &  1.08 & 107.9 &  35.0 &  0.95 & 0.001 \\
  SA1b & 21 & 1.0 & 124.7 &  35.0 & 1.08 & 104.2 &  35.0 & 0.92 & 0.168 \\
  SA1c & 30 & 1.0 & 124.7 &  35.0 & 1.08 & 101.9 &  35.0 & 0.90 & 0.219 \\
  SA1d & 45 & 1.0 & 124.7 &  35.0 & 1.09 &  96.3 &  34.3 &  0.88 & 0.261 \\
  SA1e & 60 & 1.0 & 125.1 &  35.0 & 1.03 &  90.5 &  10.4 &  0.94 & 0.069 \\
  \\[-4pt]
  SA2a & 0 & 1.4 & 123.1 &  35.0 & 1.09 & 102.8 &  35.0 &  0.93 & 0.001 \\
  SA2b & 21 & 1.4 & 122.9 &  35.0 & 1.09 &  97.2 &  35.0 &  0.89 & 0.210 \\
  SA2c & 30 & 1.4 & 123.0 &  35.0 & 1.10 &  93.8 &  34.6 &  0.87 & 0.263 \\
  SA2d & 45 & 1.4 &  99.1 &  10.8 & 1.02 &  88.5 &  10.4 &   0.93 & 0.065 \\
  SA2e & 60 & 1.4 &  99.9 &  10.2 & 1.01 &  93.2 &  10.1 & 0.95 & 0.034 \\
   \\[-4pt]
  SA3a &  0 & 3.0 &  87.9 &  35.0 &  0.84 &  80.3 &  34.7 & 0.78 & 0.001 \\
  SA3b & 21 & 3.0 &  85.5 &  11.5 & 1.00 &  67.6 &  11.1 &  0.84 & 0.070 \\
  SA3c & 30 & 3.0 &  92.6 &  10.5 & 1.03 &  74.6 &  10.3 &  0.88 & 0.064 \\
  SA3d & 45 & 3.0 &  98.2 &  10.2 & 1.03 &  85.2 &  10.0 &  0.92 & 0.048 \\
  SA3e & 60 & 3.0 &  99.8 &  10.0 & 1.01 &  92.8 &  10.0 &  0.95 & 0.021 \\
   \\[-4pt]
  SA4a &  0 & 5.0 &   0.4 &   0.3 & 0.11 &   0.1 &   0.1 &  0.07 & 0.229 \\
  SA4b & 21 & 5.0 &  69.5 &  10.3 & 0.90 &  58.0 &  10.1 & 0.79 & 0.059 \\
  SA4c & 30 & 5.0 &  85.5 &  10.2 & 1.01 &  67.8 &  10.1 &  0.85 & 0.051 \\
  SA4d & 45 & 5.0 &  96.4 &  10.1 &  1.04 &  80.7 &  10.0 &  0.91 & 0.048 \\
  SA4e & 60 & 5.0 &  99.5 &  10.0 &  1.02 &  89.9 &  10.0 &  0.94 & 0.026 \\
  FLASH \\
   FA1c & 30 & 1.0 & 121.1 & 34.4 & 12.67 & 117.1 & 34.4 &  3.34 & 0.066\\
   FA2a &   0 & 1.4 & 115.3 & 34.3  & 7.09 & 108.6 & 34.3 & 2.58 & $5\times10^{-5}$ \\ 
   FA3a &   0 & 3.0 & 8.5 & 5.4 & 6.49 & 8.5 & 5.4 & 6.49 & $10^{-4}$\\
   FA3c & 30 & 3.0 &  83.2 & 10.0 & 8.44 & 76.9 & 10.0 & 2.34 & 0.003 \\ 
   \hline
\end{tabular}
\caption{\label{tab:sph25}Collisions between a 100-$M_{\oplus}$ gas giant
  and a 25-$M_\oplus$ embryo. The gas giant planet was modelled with
  30,000 particles.  The number of particles in the impactors were
  adjusted to maintain equal particle mass with the gas giant target
  planet. At the end of the simulation the mass and radius of the material contiguous with the core of the initial gas giant planet are listed  as $M_{\rm cc}$ and $R_{\rm cc}$.  The total mass gravitationally
  bound particles and their equivalent radius are labelled  as $M_{\rm
    bm}$ and $R_{\rm bm}$.  The difference between
these quantities reflects material in a protosatellite disk or distant
material that has yet to fall back to the planet. 
All masses are in $M_{\oplus}$.
Planets' initial radius and the asymptotic radius when a quasi hydrostatic
equilibrium is re-established after the GIM's are represented by $R_{\rm T}$ and
$R_{\rm cc}$ respectively, where $R_{\rm T}=5.99\times 10^9$ cm. In the
asymptotic state, we also compare mergers' angular momentum
$J_{\rm cc}$ and compare it with that associated with rotational break up.}
\end{table*}
 
The SPH simulations show that 
head-on collisions (with $v_{\rm imp} = 1-3 v_{\rm esc,2}$) lead
to the capture of the impactor by the gas giant and the coalescence of
the impactor with the core. In addition, the envelope re-establishes a
spherical symmetry within a dynamical time scale. These results
justify the 1-D prescription and core-impactor coalescence assumption
used in our LHD scheme (see Paper I). For sufficiently high velocity
encounters, the energy dissipation near the core leads to a thermal
expansion of the envelope. In model SA3a with $v_{\rm imp} = 3 v_{\rm
esc,2}$, the impactor's kinetic energy is comparable to the gas giant's
gravitational binding energy and a significant fraction of the
envelope is lost. The requirement
\begin{equation}
v_{\rm imp} > v_{\rm cri} = (2 (M_{\rm T}+M_{\rm I})/M_{\rm I})^{1/2} 
v_{\rm esc,2}
\label{eq:breakup}
\end{equation} 
corresponds to the condition for total disruption of isolated gas
giants. Note that in model SA4a with $v_{\rm imp} = 5 v_{\rm esc,2}$,
the impact energy is so large that both the envelope and core totally
disintegrated. \citet{Anderson:2012aa} predicted a penetration velocity for head-on collisions four orders of magnitude greater than the escape velocity. Such energetic collisions likely lead to destruction of the gas giant other than penetration according to our results.

Despite the general agreement between Equation \ref{eq:breakup} and
the SPH results in Table \ref{tab:sph25}, some of the results obtained
through these SPH simulations are possibly due to numerical artefacts. For
example, the ratio of the final to initial planetary radius
($R_{\rm f}/R_{\rm i}$) remains close to unity for head-on collisions with $v_{\rm
imp} < 3 v_{\rm esc,2}$. This negligible change in the gas giant's
envelope is partially due to the artificial lower density limit we
have adopted for the SPH scheme ({\it i.e.} $\rho_{\rm l} = 0.5$ g cm$^{-3}$).
This restriction prevents
the dispersal of representative fluid elements into tenuous medium.
This technical limitation is clearly shown by the sharp density 
fall off at the gas giant's new radius. This prescription may also 
over suppress the loss of gas in the giant's original envelope as a 
consequence of high-speed collisions (see \S\ref{sec:comparison}). 
The actual critical impact speed required for substantial 
collision-induced mass loss may be substantially reduced
in the proximity of the host star. 

Without such restriction of lower density limit, the FLASH code is ideal to study the expansion of the gas giant's envelope after the impact. Consequently, the determination of different boundaries in FLASH simulations is also challenging (see Section \ref{sec:discrepancy}). In run FA2a we find much larger $R_{\rm cc}$ and $R_{\rm bm}$ than those in run SA2a, while the enclosed masses $M_{\rm cc}$ and $M_{\rm bm}$ are slightly less than that in SPH. Besides, in model FA3a with $v_{\rm imp}$ 3 times the $v_{\rm esc,2}$, the impactor is able to disintegrate the gas giant, while under the same circumstances, SPH model SA3c shows that most of the original mass of the gas giant is retained after the impact. 

We also illustrate the results of oblique collisions, first with the
parabolic impact speed ($v_{\rm imp} = v_{\rm esc,2}$). In this
low-energy limit, the impactor is always consumed by the gas giants.
Over most range of $\xi$, say $\xi < 60^{\circ}$, the impactor is able 
to penetrate through the gas giant's envelope, to merge with its core 
in a few dynamical timescales. This tendency is well illustrated in Figure \ref{fig:f1}, which
shows the SPH particles distribution of model SA1c and density slice of model FA1c in the orbital plane (designated to be the {\it X}-{\it Y} plane with the initial motion along the {\it X}-axis) for four different epochs. We also plot the particles distribution and density slice in the the {\it X}-{\it Z} plane (where {\it Z}-axis is the axis normal to the orbital plane).

We notice that the asymptotic core radius increases
with $\xi$ (in comparison with the Figure 1 in Paper I). 
In the limit of collisions with relatively large $\xi$,
the impactor is disintegrated in the envelope well outside the iron
core. If the debris of the impactor remains separated from gaseous material in the envelope of the gas giant, double-diffusive convection may kick in, and the compositional transport rate could be either very small or relatively large \citep{Rosenblum:2011aa}. In other word, whether the debris will dissolve in the envelope or eventually sediment into the core is not clear yet. 
Either way, the time scale to re-establish a hydrostatic equilibrium 
is considerably longer than the dynamical timescale. This
fractionation process provides a protracted source of energy to
sustain a relatively large planet radius $R_{\rm p}$ after the impact.
The FLASH run FA1c shows similar consequences of the impact except that the final structure is much puffier and mass loss is slightly higher, as we have noticed in the head-on cases. 

The bottom panel of Figure \ref{fig:f1} is an edge-on view of the planet after the impact, which indicates a rotational flattening structure in the gas giant which has been struck by an embryo at a modest grazing angles $30^{\circ}$. 
The capture of an embryo modifies the spin angular momentum of the gas giant by an amount $\Delta J \simeq M_{\rm I}\; R_{\rm T}\; v_{\rm imp} \sin \xi$. If the gas giant does not have any initial spin, the ratio of its final angular moment $J_{\rm f} = \Delta J$ and that corresponds to a rotational break up ($J_{\rm c} \sim (M_{\rm T} + M_{\rm I}) R_{\rm p} v_{\rm esc,2}$) would be
\begin{equation}
\frac{J_{\rm f}}{J_{\rm c}} \simeq  
   \frac{M_{\rm I}}{(M_{\rm T}+M_{\rm I})} 
   \left( \frac{v_{\rm imp}}{v_{\rm esc,2}}\right) \; \sin \xi \,,
\label{eq:angularmomentum}
\end{equation}
where we have assumed $ R_{\rm p} \simeq R_{\rm T}$.
This estimate is in good agreement with those listed in Table
\ref{tab:sph25}. In general, the angular momentum deposited by the impactor is not uniformly distributed throughout the planet. And one may notice that the ellipticity of isodensity contours of the edge-on view of run FA1c slightly decreases from inside to outside. Furthermore, different types of giant impact produce distinguished signatures of oblateness (see Section \ref{sec:oblateness}). Here we simply assume this spin angular momentum is smoothly distributed throughout the gas giant's envelope as a first-order approximation, it would speed with a frequency
$\Omega_{\rm f} \sim (J_{\rm f}/J_{\rm c}) (v_{\rm esc, 2}/\alpha_{\rm p} R_{\rm p})$, 
where $\alpha_{\rm p}$
is the coefficient for the moment of inertia.
 For isolated long-period
gas giants, a parabolic GIM by a super-Earth impactor with $M_{\rm I} < 0.1
M_{\rm T}$ would not lead to rotational break up. But, in the limit that
$J_{\rm f} = \Delta J$,
\begin{equation}
\frac{\Omega_{\rm f}}{ n_{\rm p}} \simeq \frac{M_{\rm I}}{(M_{\rm T}+ M_{\rm I})} 
\left( \frac{v_{\rm imp}}{ v_{\rm esc,2 }}\right) 
\left( \frac{3 R_{\rm H}}{R_{\rm p}} \right)^{3/2}\;  \sin \xi \,,
\end{equation}
where $R_{\rm H}$ is the Hill radius of the giant planet.
For hot Jupiters, $\Omega_{\rm f}$ can be significantly larger than the rotational
frequency associated with the spin-orbit synchronization ($\sim n_{\rm p}$). 
Thus, stochastic parabolic collisions by super-Earths can offset 
synchronous spin generally associated with star-planet tidal interaction.

For high-energy oblique collisions, impactor's capture probability
decreases with both $\xi$ and $v_{\rm imp}$. During impactor's first
passage, the total amount of kinetic energy lost due to hydrodynamic drag
alone is
\begin{equation}
\Delta E_{\rm k} = M_{\rm I}(v_{\rm imp}^2-v_{\rm final}^2)/2 \simeq M_{\rm I} v_{\rm imp} \Delta v /2 \,,
\end{equation} 
where $v_{\rm final}$ is the impactor's velocity after it passes through the target and $\Delta v = v_{\rm imp} - v_{\rm final}$ is the change of the impactor's velocity. Because momentum is conserved, we have $M_{\rm I} \Delta v \simeq M_{\rm enc} v_{\rm imp}$ and 
\begin{equation}
M_{\rm enc} (\xi) \simeq \int_L \rho A dl
\simeq {2 A R_{\rm p} \cos \xi \over 1 - \sin \xi } \int_{\sin
\xi} ^1 \rho (x) dx
\label{eq:menc}
\end{equation}
is the ``air mass'' encountered by the impactor in the gas giant's
envelope, $L$ is impactor's trajectory inside the gas giant and the
gas density $\rho$ is a function of distance from the gas giant's
center $R$ or equivalent a function of $x \equiv R/R_{\rm p}$. For modest
impacts, impactor's effective cross section $A$ is comparable to its
physical radius $R_{\rm I}$. In the above expression $M_{\rm enc}$ is a
rapidly decreasing function of the impact angle $\xi$.

We now determine the capture cross section of embryos by the gas giant.
An impactor is unlikely to be captured in the limit that $\Delta E_{\rm k} < 
M_{\rm I} (v_{\rm imp}^2 - v_{\rm esc, 2}^2)/2$ or equivalently

\begin{equation}
M_{\rm enc} < M_{\rm I} \left( 1 - (v_{\rm esc, 2}/v_{\rm imp})^2\right)/2 \,. 
\label{eq:airmass}
\end{equation}

The air mass required for capturing a intruding embryo with low impact
speed is a small fraction of $M_{\rm I}$. Provided the impact parameter is
smaller than the gas giant's radius $R_{\rm p}$ ({\it i.e.} $\cos \xi$ is 
non-negligible), most parabolic encounters leads to coalescence. But, 
the required air mass $M_{\rm enc} \sim M_{\rm I}/2$ for high impact speed
collisions. Capture of relatively massive embryos would be possible
only for relatively small $\xi$, i.e. nearly head-on collisions.
Similar arguments have been used to estimate the fraction of material lifted
into the proto-lunar disk via giant impact \citep{Canup:2008sph} and the mechanical
energy dissipated in collisions between stars \citep{Freitag:2005sc} and planets \citep{Leinhardt:2012a}.

\begin{figure}
  \centering
  \includegraphics[width= 0.98\linewidth,clip=true]{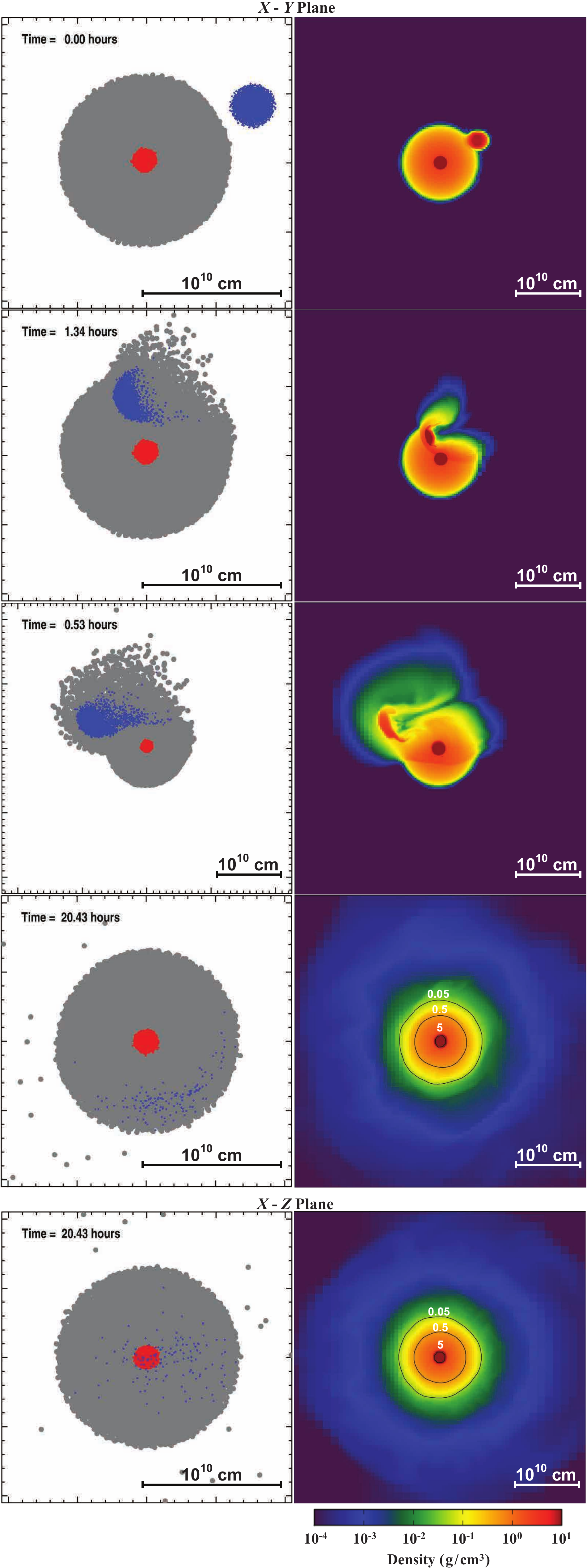}
\caption{SPH run SB3c (left) and FLASH run FB3c (right) simulations of the collision between a Saturn-like gas giant and a 10 $M_\oplus$ impactor. In the SPH simulation, gas particles are shown as grey, and core particles are over plotted in red. The basalt impactor is over plotted in blue.}
\label{fig:f2}
\end{figure}

The capture condition in equation \ref{eq:airmass} is in good
agreement with the simulation results in Table \ref{tab:sph25}, {\it
i.e.} collisions with a high impact speeds and angles generally do not
lead to capture whereas embryos which impinges onto gas giants with
relative low impact speeds and angle generally merge with them.
These general outcome morphologies have also been
observed in collisions between stars and planets
\citep{Agnor:2004ec,Freitag:2005sc,Asphaug:2006hr,Marcus:2009se,Leinhardt:2012a}
and appear a scale-invariant feature of gravity-dominated collisions.

With a high relative speed, the impactor also disintegrates when it
encounters an air mass comparable to its own. Because when $M_{\rm
enc} \sim M_{\rm I}$, the change of the impactor's total energy is comparable to its gravitational binding energy. 
Thus, in the limit of high velocity encounters,
captured embryos rapidly disintegrate whereas the flyby embryos
essentially retain their initial mass.

Finally, we point out that among all the simulations in which the planet is not destroyed, the mass loss is preferentially comes from the envelope other than the core. In fact, there is only very tiny mass loss contributed by the core. This result indicates that giant impacts may lead to the enhanced metallicity in the envelopes of gas giants.

\subsection{Collisions between a 10-$M_\oplus$ embryo and a 
Saturn-mass gas giant} 

Here, we consider a series of models which simulate the collision
between a Saturn-like giant planet and a 10-$M_\oplus$ embryo. Model
parameters and results of this series of simulations are summarized in
Table \ref{tab:sph10}. 
We emphasize here that relatively low-mass embryos
disintegrate in the envelope before reaching the core.

\begin{table*}
\centering
\begin{tabular}{cccccccccc}
   \hline\hline Model & $\xi$ & $v_{\rm imp}/v_{\rm esc,2}$  &  $M_{\rm bm}$  & $M_{\rm bm,c}$  & $R_{\rm bm}/R_{\rm T}$  & $M_{\rm cc}$ & $M_{\rm cc,c}$  & $R_{\rm cc}/R_{\rm T}$ & $J_{\rm cc}/J_{\rm *}$ \\
    \hline
    SPH \\
  SB1a &  0 & 1.0 & 110.1 &  20.0 &  1.06 &  96.9 &  20.0 &  0.95 & 0.001 \\
  SB1b &  21 & 1.0 & 110.1 &  20.0 &  1.05 &  97.0 &  20.0 &   0.95 & 0.097 \\
  SB1c &  30 & 1.0 & 110.0 &  20.0 &   1.05 &  96.2 &  20.0 &   0.94 & 0.132 \\
  SB1d & 45 & 1.0 & 109.7 &  20.0 &   1.05 &  94.3 &  18.7 &  0.93 & 0.155 \\
  SB1e &  60 & 1.0 & 109.8 &  20.0 &  1.03 & 100.2 &  19.6 &   0.96 & 0.246 \\
\\[-4pt]
  SB2a &  0 & 1.4 & 110.1 &  20.0 &   1.07 &  92.8 &  20.0 &   0.94 & 0.001 \\
  SB2b & 21 & 1.4 & 109.5 &  20.0 &   1.07 &  92.2 &  20.0 &   0.93 & 0.122 \\
  SB2c & 30 & 1.4 & 109.1 &  20.0 &   1.06 &  91.0 &  19.7 &   0.92 & 0.163 \\
  SB2d & 45 & 1.4 & 100.3 &  11.4 &   1.02 &  91.2 &  11.0 &   0.94 & 0.056 \\
  SB2e & 60 & 1.4 & 100.2 &  10.4 &   1.00 &  94.0 &  10.1 &   0.95 & 0.023 \\
\\[-4pt]
  SB3a &  0 & 3.0 & 100.5 &  20.0 &   1.06 &  79.4 &  20.0 &   0.89 & 0.002 \\
  SB3b & 21 & 3.0 &  93.3 &  11.6 &  1.04 &  73.7 &  11.1 &   0.88 & 0.070 \\
  SB3c & 30 & 3.0 &  95.6 &  10.6 &   1.04 &  79.8 &  10.3 &   0.91 & 0.061 \\
  SB3d & 45 & 3.0 &  98.7 &  10.1 &   1.02 &  87.3 &  10.0 &   0.93 & 0.038 \\
  SB3e & 60 & 3.0 &  99.9 &  10.1 &  1.01 &  93.2 &  10.0 &   0.95 & 0.017 \\
\\[-4pt]
  SB4a &  0 & 5.0 &  64.9 &  19.7 &   0.78 &  55.7 &  19.3 &   0.69 & 0.003 \\
  SB4b & 21 & 5.0 &  82.0 &  10.5 &   0.99 &  64.5 &  10.2 &   0.83 & 0.049 \\
  SB4c & 30 & 5.0 &  90.8 &  10.2 &   1.03 &  72.4 &  10.1 &   0.88 & 0.049 \\
  SB4d & 45 & 5.0 &  97.5 &  10.0 &   1.03 &  83.6 &  10.0 &   0.92 & 0.042 \\
  SB4e & 60 & 5.0 &  99.6 &  10.0 &   1.01 &  91.4 &  10.0 &  0.95 & 0.022 \\
FLASH \\
   FB3a & 0  & 3.0 & 65.9 & 14.9 & 4.52 & 65.9 & 14.9 & 4.52 & 10$^{-4}$\\
   FB3c & 30 & 3.0 & 88.9 & 10.1 &  8.42 & 80.5 & 10.0 &  1.55 & 0.004 \\ 
   FB4a & 0  & 5.0 & - & - & - & - & - & - & - \\
\hline
\end{tabular}
\caption{\label{tab:sph10}Collisions between a Saturn-mass gas giant and a 10-$M_\oplus$ embryo.
The symbols have identical meanings of those in Table \ref{tab:sph25}.}
 \end{table*}

The results in Table \ref{tab:sph10} are qualitatively similar to those in Table
\ref{tab:sph25}, {\it i.e.} all head-on and all low-speed collisions
lead to the capture of this relatively low-mass embryo and oblique
high-velocity embryos pass through and escape from the gas giant's
envelope.  

In contrast to the 25-$M_\oplus$ models, this lower mass
(10-$M_\oplus$) impactor's kinetic energy is inadequate to disrupt gas 
giant's envelope in all SPH simulations. Note that in FLASH simulations,
only the high-speed head-on collision (FB4a) can disrupt the gas giant. In some
high-speed impacts (such as model SB4b), sufficient energy is
deposited to the envelope to induces fractional mass loss, even though
the impactor is not captured during the encounter. Nevertheless, it
can carry sufficient amount of energy, especially through
high-velocity encounters, to perturb the structure of the gas giant 
(albeit SPH scheme may not be able to resolve these changes), Captured embryos can also deposit a significant amount of spin angular momentum to the gas giant.

Model parameters for the low-mass embryo to be marginally captured are
in good agreement with equation (\ref{eq:airmass}). The captured
models indicate that spin angular momentum deposited into the envelope
is greater than that needed for some hot Jupiters to uniformly
rotate with an angular frequency which synchronizes with their orbital
frequency. In contrast, gas giant's mass and angular momentum are not
significantly affected by fly-by encounters. But, in some models (such
as SB2d and SB3b) a fraction of the impactor's mass can be stripped from
it during its passage through (rather than capture by) the gas giant's
envelope. As a result, the final core mass slightly increases. Model SB2c
indicates that during oblique collisions with modest energy, this
modest size embryo is mostly disrupted, which is similar to model SA1c
except that most of the mass is deposited in the gas
giant's envelope, slightly outside the iron core. The kinetic energy
of the impactor is also deposited in the envelope rather than near the
core. Eventual sedimentation of impactor's fragments releases
additional energy on a much longer time scale (see \S\ref{sec:lhd}).
In contrast, model SB3c shows that highly energetic embryos may not be
retained by the envelope of the gas giant that they have oblique collisions
with. In FLASH simulation FB3c, the impactor penetrate through the gas giant's envelope as well, and causing over 10\% of gas giant's mass is unbound and it outer envelope becomes very extensive but its inner region remains unchanged (Figure \ref{fig:f2}).
The amount of angular momentum deposition is also
correspondingly less that the total-coalescence models with the same impact parameter.

These figures also indicate that although spherical symmetry is
temporarily destroyed during the impact, it is quickly restored both
in the envelope and the core after 1-2 dynamical time scales. Despite
the injection of spin angular momentum to the gas giant's envelope,
the asymptotic distribution of the debris is approximately spherically
symmetric.

\subsection{Merger between two Saturn-mass gas giants}

In this subsection, we present simulations of collisions between two
identical gas giants. Similar to the previous section, we adopt an
initial model with a 90-$M_\oplus$ envelope around a 10-$M_\oplus$
core. We also explore a similar series of initial conditions which
are listed in Table \ref{tab:saturn-100}. 

\begin{figure}
  \centering
  \includegraphics[width=0.98 \linewidth,clip=true]{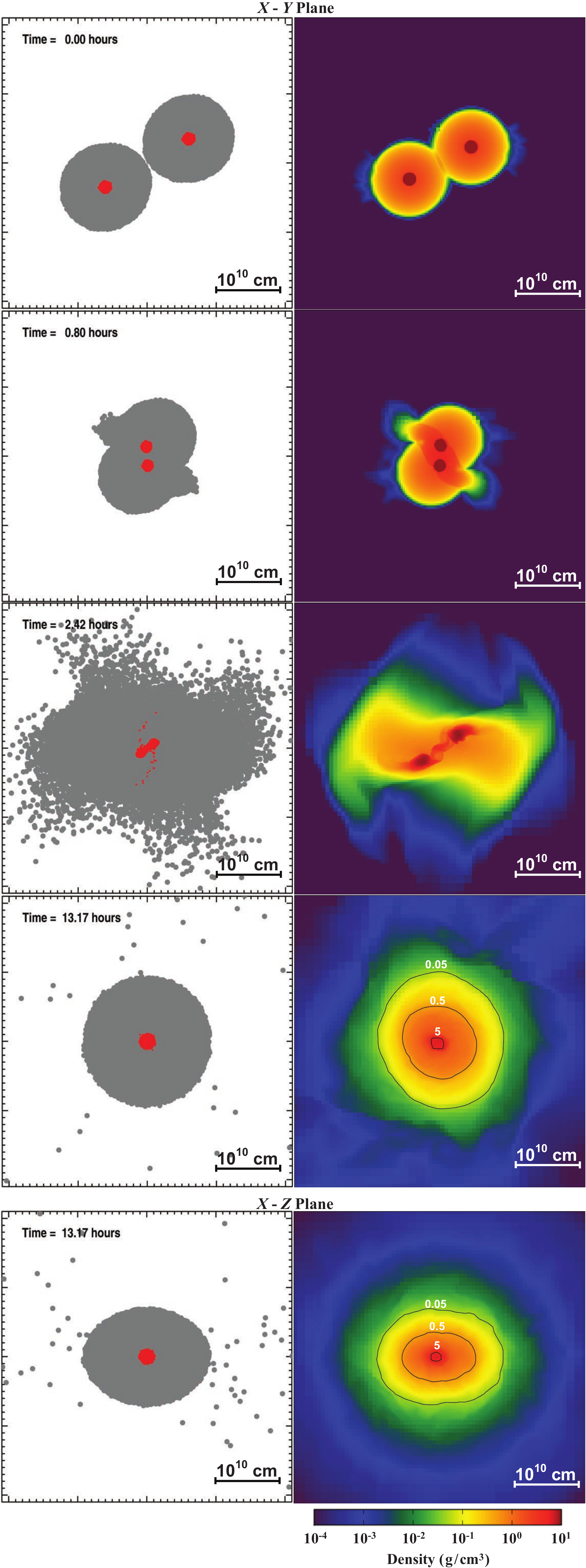}
\caption{SPH run SC1b (left) and FLASH run FC1b (right) simulations of the merger between two Saturn-like gas giants. In the SPH simulation, gas particles are shown as grey, and core particles are over plotted in red.}
\label{fig:f3}
\end{figure}

\begin{table*}
\centering
\begin{tabular}{cccccccccc}
   \hline\hline Model & $\xi$ & $v_{\rm imp}/v_{\rm esc,2}$  &  $M_{\rm bm}$  & $M_{\rm bm,c}$  & $R_{\rm bm}/R_{\rm T}$  & $M_{\rm cc}$ & $M_{\rm cc,c}$  & $R_{\rm cc}/R_{\rm T}$ & $J_{\rm cc}/J_{\rm *}$ \\
    \hline
    SPH \\
  SC1a  &   0 & 1.0 & 199.6 &  20.0 & 1.25 & 178.0 &  20.0 &  1.13 & 0.000 \\
  SC1b &  21 & 1.0 & 200.1 &  20.0 & 1.27 & 159.8 &  20.0 &  1.05 & 0.258 \\
  SC1c &  30 & 1.0 & 200.3 &  20.0 &  1.32 & 144.9 &  20.0 &  1.02 & 0.299 \\
  SC1d & 45 & 1.0 & 200.3 &  20.0 &  1.27 & 151.5 &  20.0 &  1.02 & 0.744 \\
  SC1e &  60 & 1.0 & 200.3 &  20.0 &  1.29 &  87.6 &  10.0 &   0.92 & 0.118 \\
    \\[-4pt]
  SC2a &   0 & 1.4 & 197.8 &  20.0 & 1.27 & 168.0 &  20.0 &  1.11 & 0.001 \\
  SC2b &  21 & 1.4 & 199.3 &  20.0  & 1.33 & 138.7 &  20.0  & 1.00 & 0.276 \\
  SC2c &  30 & 1.4 & 199.9 &  20.0  & 1.34 &  77.3 &  10.0 &  0.88 & 0.150 \\
  SC2d & 45 & 1.4 & 100.0 &  10.0  & 1.04 &  85.4 &  10.0 &  0.92 & 0.087 \\
  SC2e &  60 & 1.4 & 100.0 &  10.0 & 1.02 &  91.0 &  10.0 &   0.94 & 0.063 \\
    \\[-4pt]
  SC3a &   0 & 3.0 & 112.1 &  20.0 &   0.89 &  98.1 &  19.5 &  0.82 & 0.001 \\
  SC3b &  21 & 3.0 &  69.2 &  10.0 &   0.84 &  62.1 &  10.0 &   0.77 & 0.155 \\
  SC3c &  30 & 3.0 &  87.4 &  10.0 &   1.01 &  70.7 &  10.0 &   0.86 & 0.069 \\
  SC3d & 45 & 3.0 &  97.5 &  10.0 &   1.04 &  81.0 &  10.0 &  0.91 & 0.061 \\
  SC3e &  60 & 3.0 &  99.6 &  10.0 &   1.01 &  91.3 &  10.0 &   0.95 & 0.034 \\
   \\[-4pt]
  SC4a &   0 & 5.0 &  41.7 &  13.1 &   0.55 &  36.3 &  12.4 &   0.51 & 0.005 \\
  SC4b &  21 & 5.0 &  48.0 &  10.0 &   0.62 &  44.0 &  10.0 &   0.57 & 0.126 \\
  SC4c &  30 & 5.0 &  70.0 &  10.0 &   0.87 &  62.9 &  10.0 &   0.80 & 0.084 \\
  SC4d &  45 & 5.0 &  93.6 &  10.0 &   1.04 &  74.9 &  10.0 &   0.88 & 0.055 \\
  SC4e &  60 & 5.0 &  99.2 &  10.0 &   1.03 &  87.3 &  10.0 &   0.93 & 0.037 \\
 FLASH \\
  FC1b  &  21 & 1.0 & 196.3 & 20.0 & 9.76 & 188.1 & 20.0 & 3.43 & 0.181\\
  FC2a  &  0   & 1.4 & 169.2 & 20.0 & 12.47 & 148.0 & 20.0 & 3.49 & $2\times 10^{-4}$\\
  \hline
  \end{tabular}
  \caption{Collisions between two 100-$M_\oplus$ gas giant planets\label{tab:saturn-100}.
The symbols have identical meanings of those in Table \ref{tab:sph25}.}
\end{table*}

These models again indicate that parabolic encounters lead to the
total coalescence of gas giants without any loss of heavy elements in
the cores and gas in the envelopes. In order to compare with results
of the head-on parabolic merger, we present the results of model SC1b and FC1b
in Figure \ref{fig:f3}. 

These results indicate that the cores preserve their integrity until
they are in contact with each other. During the collision, the iron
cores of both gas giants are thoroughly mixed each other. These
simulations also show that the internal structure of the merger
product rapidly becomes symmetric about the Z-axis. Even with this
small impact angle, the merger product acquires a substantial amount
of spin angular momentum, primarily due to the relatively large mass
of the impactor. 

In the SPH simulation, the asymptotic major axis (in the orbital planet) of
the merger product is larger than the gas giant's initial radius $R_{\rm p}$.
The polar radius $R_{\rm po}$ (i.e. $R_{\rm p}$ in the {\it X}-{\it Z} plane) remains unaltered from the initial $R_{\rm p}$. But this difference between major and minor axes
indicate that the enlargement of equatorial radius $R_{\rm eq}$ (i.e. $R_{\rm p}$ in the {\it X}-{\it Y} plane) is mainly due to
the rotational effects rather than thermal expansion. Indeed, the
amount of energy dissipation during the collision is actually a modest
fraction of the gas giant's initial internal energy. 

In the FLASH simulation, both $R_{\rm po}$ and $R_{\rm eq}$ of the merger product expand, but $R_{\rm eq}$ is also much larger than $R_{\rm po}$. We will discuss the oblateness of FLASH simulations in Section \ref{sec:oblateness}.

For highly oblique collisions (models SC1d and SC1e), the spin angular
momentum of the merger product is comparable to the break up
speed. There is likely to be substantial tidal evolution toward spin
synchronization. The continuous influx of internal energy can lead to
substantial tidal heating and possibly loss of envelope mass \citep{Gu:2003ez,
Gu:2004km}. 

For relatively high impact velocities ($v_{\rm imp}/v_{\rm esc,2} =1.4$) and
small-impact angle collisions ($\xi<30^{\circ}$), the condensed
materials of the impactor and target effectively merge with little
loss of the gaseous envelope. However, for more oblique collisions
(models SC2d and SC2e), the energy dissipation during the impact is
insufficient to bind the colliding gas giants. They continue their
original course with a small amount of energy loss and no mass loss.

With a higher impact speed ($v_{\rm imp}/v_{\rm esc, 2} =3$), a head-on 
collision leads to the loss of most (more than half), but not all of initial
envelope gas while the two cores merged (model SC3a). For collisions with 
modest impact angles ($\xi = 21^{\circ}$, model SC3b), merger of the cores 
is avoided while each gas giant loses a substantial (up to 1/3)
fraction of its initial envelope. Given that half of collisions have
non negligible $\xi$, these non-accretionary, envelope-eroding
collisions are as probable as core-merging collisions. These 
intermediate-mass (between Jupiter and the Earth) merger products 
may account for the unexpected presence of some planets in the 
domain of ``planet desert'' \citep{Ida:2004ko, Howard:2010bs, 
Ida:2013fv}. 

For highly oblique encounters (with $\xi \ge 30^{\circ}$, models 
SC3c, SC3d and SC3e), gas giants retain their original envelope, 
albeit considerable angular momentum is deposited to the gas giant's 
spin during gas giants' brief passage
through each other's envelope. 

Hyper-velocity head-on collisions (with $v_{\rm imp}/v_{\rm esc, 2} =5$) lead to
the non-equal fragmentation of both core and envelope. Finally, after
high-velocity oblique collisions, cores continue their original fly
paths with a fraction of their initial envelopes. In contrast to the
substantial (but not entire) loses of gas giants' initial envelope,
most of the core material is retained by the merger product. Thus,
the metallicity of the merger product always increases from that of
the original gas giant planets. Finally, these collisions deposit not
only energy and heavy elements but also angular momentum to the spin
of the colliding gas giants.

\subsection{Comparison between SPH and FLASH results}
\label{sec:discrepancy}
Both SPH and grid codes have been widely used to perform hydrodynamic simulations. Because of their inherent differences, SPH and grid methods are complementary to study the same problem. For example \citet{Canup:2013aa} compared these two methods in the context of the lunar-forming impacts, and they found a general agreement between the SPH and the grid code. In this work, we confirm that
the SPH and the FLASH codes produce overall agreement on the critical impact speed for the envelope and core mass losses, even though they differ in the size of resultant collisions. 

Apart from the hydrodynamics, gravity and EOS, the two codes also differ in the treatment of the ambient medium. 
The SPH simulations incorporate a density limit $\rho_{\rm l} = 0.5$  g cm$^{-3}$ to prevent spreading of very low density gas (that is not well-resolved in an SPH model with equal particle mass). The density limit acts as an external pressure, and the expansion of the particle stops at the density limit and the particle then evolves due to gravity and whatever neighbors it is interacting with hydrodynamically.

It is known that AMR codes are very suitable to deal with surrounding structure around a dense object given that the minimum-grid and refinement criteria are made to capture the dense region \citep{Tasker:2008aa}. In our grid scheme, we adopt a density $\rho = 10^{-19}$  g cm$^{-3}$ for the ambient medium, so gas can diffuse into tenuous interplanetary space. As a result, all the FLASH simulations produce much puffier gas giants than those observed in the SPH runs.

However, to determine the corresponding contiguous and bound radii $R_{\rm cc}$ and $R_{\rm bm}$ in FLASH is less straightforward because the gas density gradually decreases all the way down to the ambient medium level at a fairly large distance (usually two orders of magnitude larger than the original size of the planet). Furthermore, simply prolong the simulation until the planet is fully relaxed is inefficient and may introduce other accumulative  errors. 

\begin{figure*}
  \centering
  \includegraphics[width= \linewidth,clip=true]{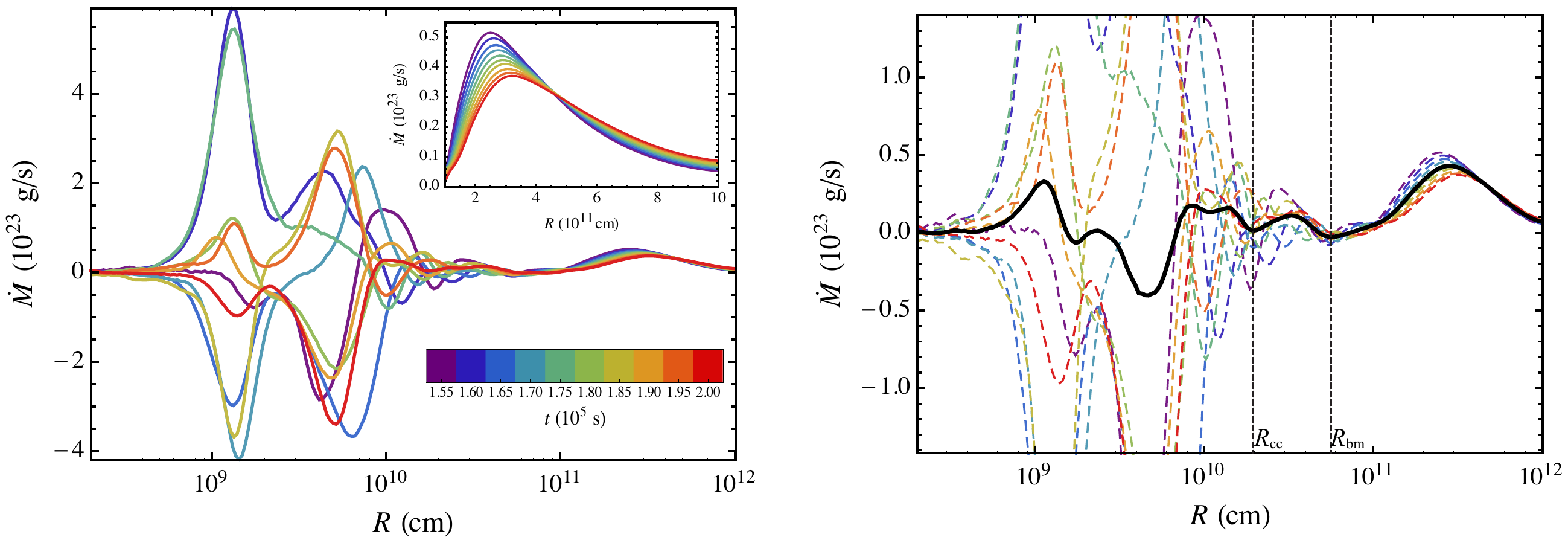}
\caption{Snapshots of mass flux \textit{\.{M}} as a function of radius taken from the FLASH simulation FC1b. In the left panel, different snapshots are plotted in colored lines. We zoom in the outflow region in the subplot. In the right panel, we plot the time-averaged \textit{\.{M}} in black thick line and the colored dashed lines are the snapshots. Two vertical black dashed lines indicate the positions of contiguous radius $R_{\rm cc}$ and bound radius $R_{\rm bm}$, respectively.}
\label{fig:f5}
\end{figure*}

Here we use the mass flux \textit{\.{M}} to quantitatively analyze the structure of the planet. The \textit{\.{M}} at a given radius is the amount of mass that is flowing out through the sphere with that radius, e.g., a positive \textit{\.{M}} indicates an outflow (and vice versa). A post-impact sequence of the mass flux \textit{\.{M}} as a function of the radius taken from FC1b is plotted in the left panel of Figure \ref{fig:f5}. Ten successive snapshots of mass flux  \textit{\.{M}} with a constant separation of 5000 s are presented in different colors. Even almost 2 days after the impact, the FLASH result shows that large-amplitude oscillations of mass flow are still prevailing throughout the whole planet, which suggests that the relaxation time scale is much longer than the dynamical timescale. On the outskirts of the the planet where the density drops below $10^{-2}$ g cm$^{-3}$ (see Figure \ref{fig:f3}), the amplitude of such oscillations gradually decreases. As the distance goes further, the amount of mass flux overturns representing the gas outflow due to the merger. The position of the peak increases with time whereas the value of the peak decrease, which suggests the inner part does not replenish the outflow. 

We plotted the time-averaged mass flux of the ten snapshots in the thick black line in the right panel of Figure \ref{fig:f5}. We define the contiguous radius $R_{\rm cc}$ as the radius within which large-amplitude oscillations of \textit{\.{M}} occur but the the averaged \textit{\.{M}} is much reduced, so the net mass flow is orders of magnitude smaller than the instantaneous value. The bound radius $R_{\rm bm}$ is defined as the separation between the bound mass and unbound mass. Material in the region surrounded by spheres with $R_{\rm cc}$ and $R_{\rm bm}$ is very tenuous and but still bound to the planet. It may form a protosatellite disk or be stripped off due to other perturbations such as tidal interactions. 

The definitions of $R_{\rm cc}$ and $R_{\rm bm}$ are somewhat less stringent in the sense that these boundaries may evolve with time. Considering the planet will cool off due to thermal radiation, the contiguous radius $R_{\rm cc}$ is is likely to become smaller. As the planet adjusts its internal structure to cooling, it becomes more similar to that produced by SPH simulations. On the other hand, the removal of the outflow may also lead to the expansion of $R_{\rm bm}$. In the context of the planetary system, the $R_{\rm bm}$ is comparable to its Hill radius. However, we notice that the net mass flow is negligible - in other words, the enclosed mass does not change despite the evolution the $R_{\rm cc}$ and $R_{\rm bm}$. In this work, we measure the two radii at the time about 2 days after the impacts or mergers in all FLASH simulations.

\subsection{Oblateness as an observable signature of GIM}
\label{sec:oblateness}
\begin{figure}
  \centering
  \includegraphics[width= \linewidth,clip=true]{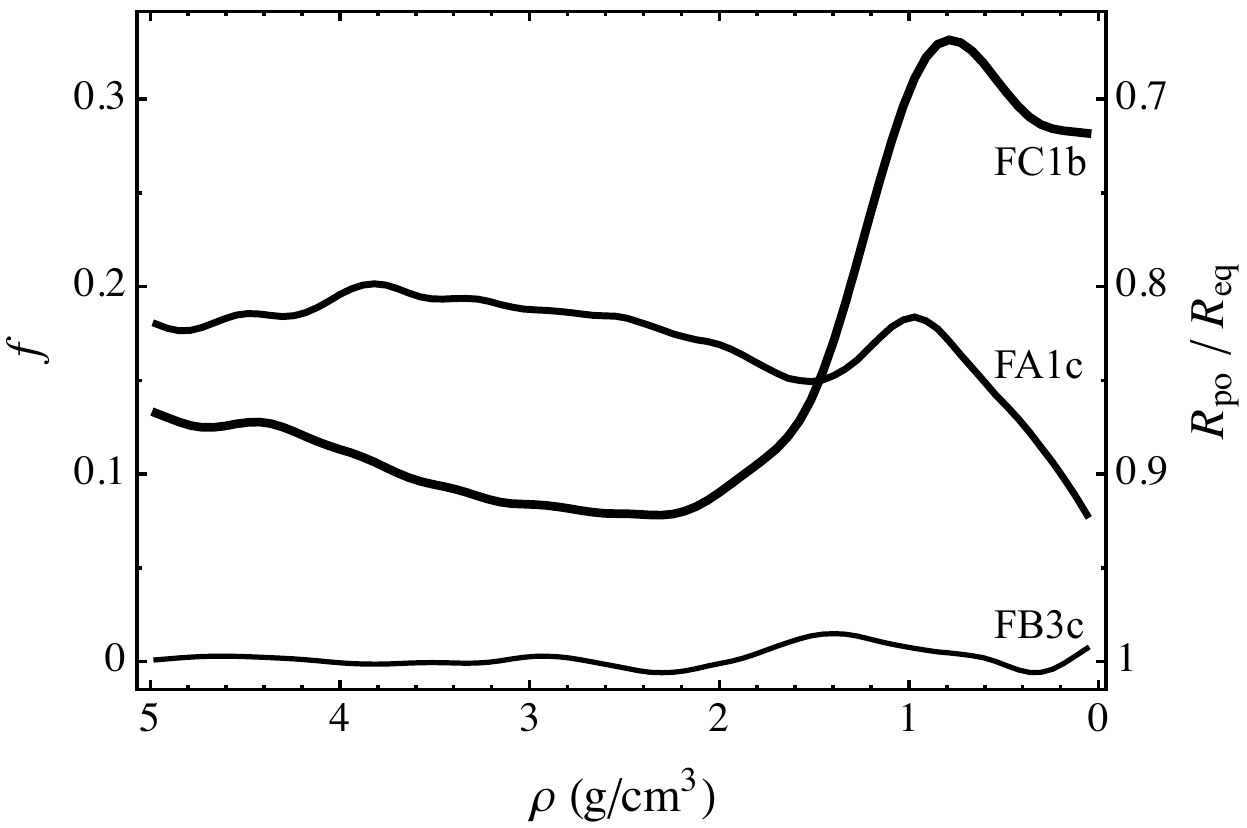}
\caption{Oblateness $f$ of isodensity surfaces in three FLASH simulations FA1c, FB3c and FC1b. The right-y axis is linearly scaled to show the corresponding $R_{\rm po}/R_{\rm eq}$ value.}
\label{fig:f6}
\end{figure}

Oblique impacts or mergers deposit angular momentum into the gas giant planet and alter its shape as well as its spin axis. To describe the shape of a planet, it is useful to define the oblateness as
\begin{equation}
f \equiv \frac{R_{\rm eq}-R_{\rm po}}{R_{\rm eq}}, 
\end{equation}
where $R_{\rm eq}$ and $R_{\rm po}$ are the equatorial radius and the polar radius, respectively. In Figure \ref{fig:f6}, we plot the oblateness $f$ of isodensity surfaces from the interior to the surface of the planet. Three FLASH simulations FA1c, FB3c and FC1b are investigated. The oblateness profile is obtained by measuring the $R_{\rm eq}$ and $R_{\rm po}$ of many isodensity surfaces. Each simulation represents a possible scenario in giant impacts or mergers.

In the simulation FB3c, a 10-$M_\oplus$ embryo penetrates the gas giant's envelope rapidly leaving the bulk structure of the gas giant mostly unchanged. The time for the impactor to exert a torque to spin up the gas giant is short. As a result, the embryo does not deposit much angular momentum into the gas giant, so the oblateness $f$ remains close to 0 throughout. 

In the other giant impact simulation FA1c, a 25-$M_\oplus$ embryo is disintegrated near the gas giant's core and deposit a large amount of angular momentum there. Therefore, the oblateness $f$ keeps a relatively high level in the interior (say $\rho>1$ g cm$^3$) and drops quickly to a sub-Saturn level at larger radii. In another words, the gas giant gains much more angular momentum than that in the previous case, but its observable shape does not differ very much from a sphere as $f$ is small in the surface layer.  

Nevertheless, the simulation FC1b shows an opposite pattern of $f$, in which $f$ is small in the interior and becomes prominent near the surface. That is because during the oblique merger of two equal mass gas giants, torque exerted on different parts varies a lot, which introduces differential rotation to the planet. As a result, the envelope spins faster than the core and $f$ peaks near the surface and remains significant at the outer boundary.

Inspired by these dramatic differences, we propose that the oblateness may serve as a proxy for probing the impact history of gas giant planets given that the exchange of angular momentum between the planet and host star can be ignored. Giant impacts generally do not lead to large oblateness, while gas giant mergers do. An oblate transiting extrasolar planet would have a discernible effect on transit light curves \citep{Hui:2002aa,Seager:2002aa,Barnes:2003aa}. Combing the oblateness information with other physical properties like mass, radius and metallicity, we may be able to constrain the formation history of a giant planet.

\section{LHD simulations of Giant Impacts and Mergers of the Gas Giants}
\label{sec:lhd}
Hydrodynamic simulations of GIM can only last for a few dynamic timescales. After the establishment of hydrostatic equilibrium, the giant planet undergoes thermal evolution on a timescale much longer than the dynamical timescale. Here we following the
long-term thermal evolution of the remnant gas giant using a
one-dimensional Lagrangian Hydrodynamic (LHD) model (see Paper I for
model details).  Combinations of SPH impact models and 1-D
long-term thermal models have been used to examine the luminosity and
core-accretion of gas-giants \citep{Broeg:2012ca} and long-term
luminosity of gas giants \citep{Anic:2007gg}.
The details of our one-dimensional LHD model are described in paper
I.  Here we use similar initial models ($M_{\rm
tot}=100~M_{\oplus}$, $M_{\rm core}=10~M_{\oplus}$, $R_{\rm tot}=7\times 10^9$
cm, $R_{\rm core}=2.2~R_{\oplus}$), but consider impacts into 
short-period planets and calculate models with a much larger range of $v_{\rm
imp}$. In order to take into account the host stars' tidal potential, we adopt
a prescription for the gravitation potential (see \S3 in Paper I). In most LHD
models presented here, we set the semi major axis of the planet to be at $a =
0.042$ AU and calculate $R_{\rm R}$ ($\sim 3 \times 10^{10}$ cm = 4.3 $R_{\rm J}$) with
$M_\ast$ being that of HD\,149026 ({\it i.e.} 1.3 $M_\odot$). We take the
stellar irradiation into account by assuming the surface temperature of the
planet to be 1500 K. For comparison, we also simulate other models (LB3e,
LA3a, LA3b, LC2a, LC2b) in which, we place the gas giant at $a=0.1$ and 5 AU
respectively. The planet's surface temperature is changed accordingly.

\subsection{LHD models of GIMs on to hot Jupiters}

In this series of simulations, we adopt the same model
parameters which we used to simulate head-on collisions with the SPH method.
The results of
9 sets of simulations are summarized in Table~\ref{tab:LHDGIM}. They
are three sets of models in which the impactor is a Saturn-mass gas
giant, a 25 $M_\oplus$ and a 10 $M_\oplus$ embryo. All
of the initial models are assumed to be in hydrostatic equilibrium prior to
the impact. At the onset of the simulation, in order to take into account
the impact energy released in the vicinity of the core,
we impose a burst of thermal energy, $E_{\rm a} = M_{\rm I} v_{\rm imp}^2/2$,
so that the total energy becomes $W^\prime = E_{\rm a} + E_{\rm g}/2$.
We also adopt the assumption that
cores and embryos are completely merged during head-on impacts. This
assumption is supported by the results of SPH models SA1a-SA3a,
SB1a-SB4a, and SC1a-SC4a.

For an impactor with $M_{\rm I}=25~M_{\oplus}$, a parabolic collision 
(model LA1) leads
to a merger without any loss of gaseous envelope. The gas giant's
photospheric radius expanded slightly. For a higher-speed collision
(model LA2), the envelope expands by a larger factor. Although it
is now comparable to some of the inflated transiting planets, this gas
giant will contract as it loses its internal energy on a Kelvin
Helmholtz time scale. For a hyperbolic encounter (model LA3), the
total energy $E_{\rm a}$ deposited into the gas giant is comparable to
its total gravitational potential energy $E_{\rm g}$ prior to the collision.
With a positive total energy after the impact ($W^\prime >0$),
the gas giant's envelope expands and disintegrates rapidly.

Analogues dynamical parameters are imposed for a 10 $M_\oplus$
impactor. In these models, the final $R_{\rm f}$ is slightly smaller for the same
impact velocity compared with those models impacted by a larger impactor (25
$M_{\oplus}$). Still, we find total loss of the gaseous envelope in the model
(LB3) with high velocity impact. 

For the gas giant merger models, LHD simulations yield greater
loss of the gaseous envelope. In the low-velocity parabolic model LC1, the
merger produces a much larger photospheric radius than the gas
giants' initial size. Nevertheless, the entire envelope is retained
due to a comparable increase in the merger's thermal and gravitational
binding energy. The new Roche lobe of the merger is also enlarged
so that $R_{\rm f}$ remains well inside $R_{\rm R}$. However, with a slightly
larger impact speed (model LC2), the thermal energy dissipated is
adequate to enlarge $R_{\rm f}$ beyond the new $R_{\rm R}$. Outflow across
$R_{\rm R}$ leads to runaway mass loss.

The results in this subsection indicate a bimodal distribution
of GIM outcomes. With high impact speeds, gas giants lose
their envelope entirely. Shortly after modest-speeds, gas giants
retain their enlarged envelopes. It is possible that subsequent
tidal evolution can lead to additional heating and modest mass
loss \citep{Gu:2003ez, Gu:2004km}. This process can account for the
smaller observed masses for hot Jupiters versus long-period gas
giants. But it cannot produce intermediate-mass (with $20 
M_\oplus < M_{\rm p} < 100 M_\oplus$) hot Jupiters in a prolific manner.
Thus, the bimodal mass distribution of planets anticipated by
planetary synthesis simulations \citep{Ida:2008dq, Ida:2013fv} is likely
to be preserved despite the high probability of GIMs in
the stellar proximity (see \S\ref{sec:justify}).

\begin{table}
\centering
\begin{tabular}{ccccccccc}
\hline\hline 
Model & $M_{\rm I}$ &$M_{\rm I,c}$  & $v_{\rm imp}/v_{\rm esc,2}$ &   $M_{\rm f,c}$  &  $M_{\rm f,g}$ & $M_{\rm f,g}^{\rm S}$ & $R_{\rm f}/R_{\rm i}$  & $R_{\rm f}/R_{\rm S}$ \\
\hline 
 LA1  &   25  & 25 &  1.0  & 35  &   90 & 90  & 1.08 & 1.25   \\
 LA2  &   25  & 25 &  1.4  & 35  &   90 & 88  & 1.22 & 1.42   \\
 LA3  &   25  & 25 &  3.0  & 35  &  0   & 53  & -    & -      \\
 LB1  &   10  & 10 &  1.0  & 20  &   90 & 90  & 1.10 & 1.28   \\
 LB2  &   10  & 10 &  1.4  & 20  &   90 & 90  & 1.12 & 1.30   \\
 LB3  &   10  & 10 &  3.0  & 20  &  0   & 81  & -    & -      \\
 LC1  &  100  & 10 &  1.0  & 20  &  180 & 180 & 1.94 & 2.25   \\
 LC2  &  100  & 10 &  1.4  & 20  &  0   & 180 & -    & -      \\
 LC3  &  100  & 10 &  3.0  & 20  &  0   & 92  & -    & -    \\
 \hline
 \end{tabular}  
\caption{LHD models of collisions of a 100 $M_\oplus$ hot Jupiter with
a 25 $M_\oplus$ embryo, a 10 $M_\oplus$ embryo, and another identical
gas giant\label{tab:LHDGIM}.
Each gas giant has a 10 $M_\oplus$ core. In all of these models,
the semi major axis of the hot
Jupiter is set to be 0.04 AU. All the symbols are identical to previous
tables. For each model, we list the asymptotic mass $M_{f,g}^S$ of
remaining gas envelope of the merger product obtained with the SPH
simulations (see Tables \ref{tab:sph25}, \ref{tab:sph10}, and
\ref{tab:saturn-100}).}
\end{table}

\subsection{Comparisons between the SPH and LHD models}
\label{sec:comparison} 

Similar to the results obtained from LHD calculations, the SPH models
in \S\ref{sec:sph} clearly demonstrate a
sharp transition between GIMs without much mass loss and total
disruption. Modest events hardly modify gas giant's radius whereas
highly energetic GIMs can lead to their total dispersal. 
However, for some models, LHD calculations generate more significant mass
loss after the impact compare with SPH results.
This may be caused by the artefact of a density lower limit ($\rho_l = 0.5$
g cm$^{-3}$) which we have adopted for the SPH scheme (see Paper I).
While the LHD scheme uses a freely expanding outer boundary condition.

In this subsection, we carry out a series of models to
demonstrate how the mass loss rate may depend sensitively on the outer
boundary condition (see Table \ref{tab:LHDE3}). We choose the
parameters for an intermediate model SB3a to highlight the difference
between SPH and LHD simulations in marginal cases. In the SPH
simulations, nearly 90\% of the envelope's original mass was retained
after a Saturnian gas giant was impacted by a 10 $M_\oplus$ embryo
during a head-on high-speed collision.

\begin{table}
\centering
\begin{tabular}{cccccccc}
\hline\hline
Model & $v_{\rm imp}/v_{\rm esc, 2}$  & $a$ & $\rho_\infty$  &  $M_{\rm f,c}$  & $M_{\rm f,g}$  & $R_{\rm f}/R_{\rm i}$  & $R_{\rm f}/R_{\rm S}$ \\
    & & (AU) &  g cm $^{-3}$ & ($M_\oplus$) & ($M_\oplus$) & & \\
   \hline 
SB3a &  3  & ---   & 0.5              &   20  &   80.5 & 1.06  & 1.05    \\
LB3  &  3  & 0.042 & free             &   20  &   0    & -    & -     \\
LB3c &  3  & 0.042 & $2 \times 10^{-5}$ & 20  &   0    & -    & -     \\
LB3d &  3  & 0.042 & $5 \times 10^{-3}$ & 20  &   0    & -    & -     \\
LB3e &  3  & 0.042 & $5 \times 10^{-2}$ & 20  &   90   & 2.80 & 3.25  \\
LB3a &  3  & 0.1   & free             & 20    &   90   & 2.38 & 2.61  \\
\hline
\end{tabular}
  \caption{Comparison of LHD models with a SPH model of collisions
between a 100 $M_\oplus$ gas giant planet and a 10 $M_\oplus$ embryo\label{tab:LHDE3}.
 All models in this table are computed for $v_{\rm imp} = 3 v_{\rm esc, 2}$.
In the SPH model, we consider the gas giant in isolation. In LHD models,
tidal effect of a 1.3 $M_\ast$ host star is taken into account.
For the LHD models, various outer
boundary conditions are imposed to demonstrate the numerical artefacts
introduced by the lower limit on the external density assumed in the
SPH scheme. For the LHD models, the final mass of envelope is computed
to be that inside the Roche radius $R_{\rm R}$ and the quantity $R_{\rm f}$ refers
to the photospheric radius when the post-impact expansion of the
planetary envelope is stalled.}
\end{table}

\begin{figure*}
\begin{center}
 \includegraphics[width= \linewidth,clip=true]{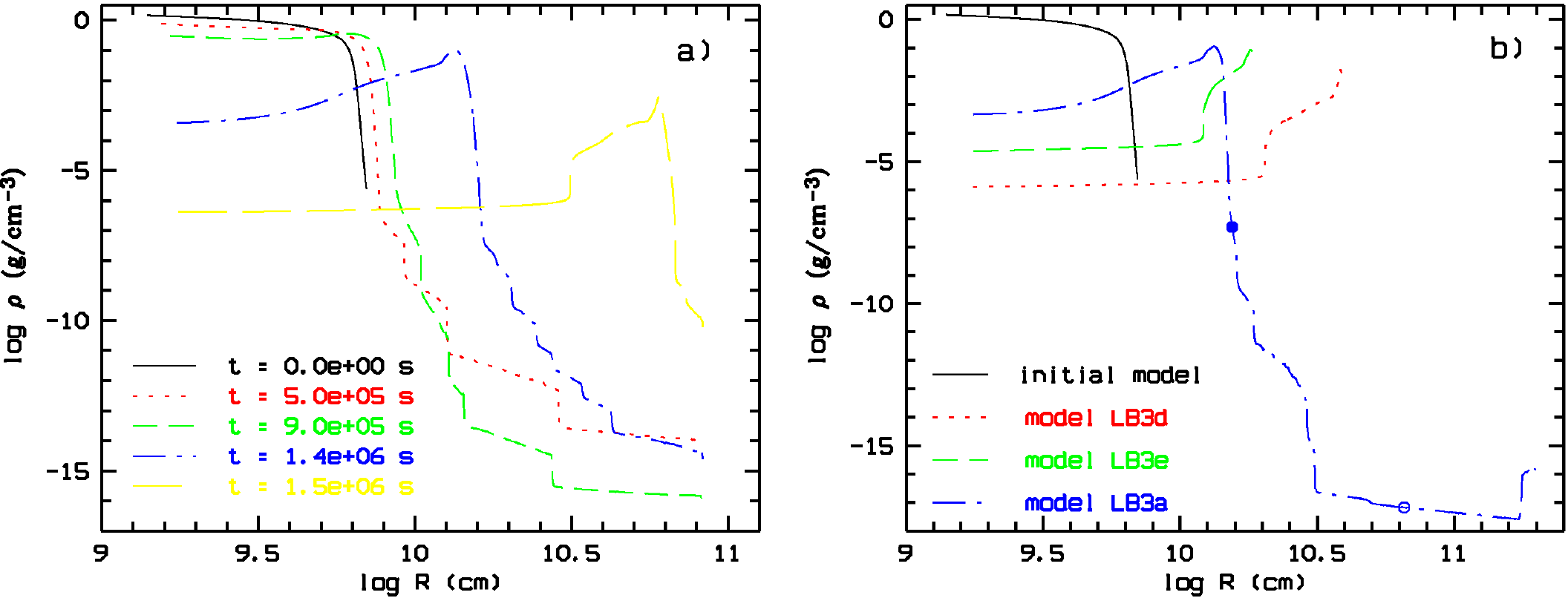}
\end{center}
\caption{Models' density profiles. Left panel: Evolution of the density
distribution at different epochs for the model LB3. Right panel: Comparison
of the density distributions for models with different boundary
conditions.}
\label{fig:LB3den}
\end{figure*}

LHD simulations suggest there may be substantial mass loss associated with this
set of parameters. In model LB3 (and most models), the pressure at outer
boundary of the planet has a smooth transition from the planetary surface to
the disk.  
The evolution of the density distribution $\rho$ (at four epochs)
after the impact is plotted on the left panel of Figure \ref{fig:LB3den}. Due
to the dissipation of impactor's kinetic energy, thermal energy is released
near the core. A jump in the pressure drives a rapid initial expansion of the
envelope (at around $5 \times 10^5$ s). Due to the $P{\rm d}V$ work, the expansion
speed gradually decreases below the sound speed. Nevertheless, gas in the
envelope is able to reach $R_{\rm R}$ (where the gravitational potential has a local
maximum) with a modest density ($\rho_{\rm R} \sim 10^{-4}$ g cm$^{-3}$) at $\sim
10^6$ s after the impact. Gas outside $R_R$ is accelerated outward by the tidal
force of the central star with a flux ${\dot M}_{\rm out} \sim 4 \pi \rho_{\rm R}
R_{\rm R}^2 v_{\rm g} > 10^{23}$ g s$^{-1}$. Gas depletion just beyond $R_{\rm R}$ reduces the
local pressure and a negative pressure gradient drives an outward gas flow to
replenish that region. Since $R_{\rm R}$ is proportional to $M_{\rm p} (R_{\rm R})$, mass loss
also decreases the size of $R_{\rm R}$. A continuous flow across $R_{\rm R}$ leads to the
eventual loss of the entire envelope.

There are some differences in the results in models LB3 and
SB3a. Although the same model parameters are used, these two models
are computed with LHD and SPH methods respectively. In order to
account for these differences, we modify the outer
boundary condition such that the planet is assumed to be surrounded by
an external medium with a density $\rho_\infty$ and a temperature $T
=1,500$ K. We impose a passive external density
$\rho_\infty$ which is set to be $2 \times 10^{-5}$, $5 \times
10^{-3}$, $5 \times 10^{-2}$ g cm$^{-3}$ for models LB3c, d, and e
respectively (This external material does not accumulate beyond the
planet as it expands). The results in Table \ref{tab:LHDE3} indicate
that, with sufficiently high $\rho_\infty$, expansion of the envelope
can be stalled before it reaches $R_R$. On the right panel of Figure
\ref{fig:LB3den}, we plot the initial and stalled density distribution
for models LB3d and LB3e.
In both models LB3d and LB3e, the external density (or
pressure) has stalled the expansion of the envelope.
The main difference is that in model LB3d, the
envelope continues to expand beyond $R_{\rm R}$ whereas in model LB3e, it is
stalled and falling back inside $R_{\rm R}$. Due to the outward-directed
tidal force, gas outside $R_{\rm R}$ does not return to the gas giant. In
contrast, the build-up of a dense shell just inside $R_{\rm R}$ (in model
LB3e) will settle back to the core region. However, the time scale
for the dense shell to resettle onto a point mass potential may be
larger than the local dynamical time scale, because the magnitude of
gravity near $R_{\rm R}$ is reduced by the host star's tidal force. The
approximate agreements suggest that our adopted value of $\rho _{\rm l}$ in
the SPH simulations can suppress the effect of envelope's mass loss,
especially for marginally disruptive models.

In all SPH models, $a$ and $R_{\rm R}$ are set to be infinite. It is 
therefore not unfavorable for any gas giant to lose a substantial
fraction of its envelope unless GIM's occur with a kinetic energy
which at least exceeds
its initial gravitational binding energy. Some of the hyper-velocity models do
have adequate kinetic energy to disintegrate the gas giant planets. Despite the
detachment and high velocity of disposal of a part (but not all) of gas in the
original envelopes, a residual amount of gas falls back to re-establish a much
reduced envelope (see models SC4a-e). The LHD calculations take into
account the stellar tide. When the gaseous envelope of the planet expands
beyond its $R_{\rm R}$ after the GIM, the gas outside $R_{\rm R}$ would not return back
to the planetary surface due to the stellar gravity. Therefore, the change
of planets' $a$ (and accordingly $R_{\rm R}$) should have influences on the
retention efficiency of the planetary atmosphere after the impact.
We calculate marginal models with different $a$'s and $R_{\rm R}$'s and 
present comparisons in the subsection below.

\subsection{GIM's on to intermediate and long-period gas giants}
\label{sec:weakentide}

We have shown in our LHD models that in the stellar proximity where hot
Jupiters' $R_{\rm R}$ is only a few times Jovian radii, energetic GIM's lead to
catastrophic loss of their gaseous envelopes. During some GIM's (such as those
in models LA3 and LB3), although the initial expansion of a gas giant's gaseous
envelope's is stalled with its $R_{\rm f}$ interior to $R_{\rm R}$, there is sufficient
density at $R_{\rm R}$ to enable a high outflow flux. In these marginal cases, it is
natural to explore whether a fraction of the initial envelope may be retained
in the limit of larger $R_{\rm R}$'s. Take model LB3 for example, the total kinetic
energy carried by the impactor is comparable to half of the total gravitational
binding energy of the pre-impact gas giant. In the limit of negligible stellar
tide, we anticipate that the planet's envelope may double its half mass radius
and increases the photospheric radius to even larger extent.

The magnitude of $R_{\rm R}$ increases with $a$. Since gas giants probably formed
outside the snow line at a few AU's from their host stars \citep{Ida:2004ko},
their $R_{\rm R}$ is two orders of magnitude larger than that of the hot Jupiters.
In order to explore dependence of the envelope retention on the tidal
perturbation of the host star, we consider variations of models LA3 and LB3 in
which $a$ is chosen to be 0.1 AU and 5 AU. The surface temperature of these
models are set to be 1000 K and 150 K accordingly. We list the models
parameters and results in Table~\ref{tab:LHDCOMP}.

In the choice of these model parameters, the target Saturn-mass gas
giant had $v_{\rm esc} = 25$ km s$^{-1}$ while the Keplerian speeds at 0.1
and 5 AU around HD\,149026 are $>100$ km s$^{-1}$ and $\sim 15$ km
s$^{-1}$ respectively. Although it is possible for GIMs to occur with
$v_{\rm imp} = 3 v_{\rm esc, 2}$ at 0.1 AU (models LA3a and LB3a), it is highly
unlikely for $v_{\rm imp}$ to be much larger than $v_{\rm esc, 2}$ at 5 AU (in
models LA3b and LB3b). Therefore, the simulation of highly energetic
GIM's onto long-period exoplanets should be considered theoretical toy
models albeit they can also be used to represent GIM's by more massive
super-Earths with somewhat smaller $v_{\rm imp}'$s.

The total energy deposited during a hyper-velocity impact in the hot-Jupiter
model LB3 is around half of the gas giant's initial gravitational binding
energy or comparable to its total energy. In principle, gas giant's envelope
should be more than double after it is fully
virialized, which is consistent with our FLASH simulations. However, $R_{\rm R}$ for
this hot Jupiter is only five times $R_{\rm p}$ prior to the impact. The initial
expansion overshot causes sufficiently dense ($\rho > 10^{-5}$g cm$^{-3}$)
outer region of the envelope to extend outside the gas giant's Roche lobe such
that it leads to rapid mass loss. In model LB3a, however, both $a$ and $R_{\rm R}$
are set to be more than twice as large as their values in model LA3.
Consequently, the envelope expansion is stalled well inside $R_{\rm R}$ with very
little mass loss. At later stages, the envelope contracts slightly as a quasi
hydrostatic equilibrium is re-established while the radius of the photosphere
is doubled from its pre-impact value. For comparison, we overplotted the
stalled density profile of this model on the right panel of Figure
\ref{fig:LB3den}. In order to demonstrate the locations of planetary
photosphere and Roche radius after the GIM, we marked them with filled and open
circles respectively. We also consider a model located at larger value of $a$
(= 5 AU). In this case, the gaseous envelope is still well preserved. 

For the models LAa and LCa, the Roche lobe is still sufficiently compact, that
there is considerable outflow and depletion of the gas envelope. In long-period
models (LAb and LCb), the $R_{\rm R}$'s are so large that the expanding envelopes can
hardly reach them. Even though, the gaseous envelopes in these models are
dispersed as their expanding velocities have already exceeded the escape
velocity.

The total thermal energy input $E_{\rm a}$ (or kinetic energy dissipation) at the
onset of these two models is comparable to the initial gravitational binding
energy of the gas giant $E_{\rm g}$. Consequently, the net total energy ($W^\prime$)
is slightly positive. During the initial expansion, this newly added
internal energy is converted into a self similar outflow everywhere analogous
to the Sedov-Taylor solution for supernova explosions. In the absence of energy
losses, the magnitude of the gravitational binding energy would reduce with
that of the internal energy while there would be adequate kinetic energy to
disintegrate the entire envelope with an outflow velocity $\sim (2 W^\prime/
M_{\rm p}) ^{1/2}$.

In the limit that sufficient thermal energy is deposited into these long-period
gas giants, it is possible for a fraction of the envelope be dispersed while
the rest is retained around the merged core. During the expansion of the
gaseous envelope after an energetic impact, a fraction of the envelope remains
around the core, albeit the asymptotic photosphere $R_{\rm f}$ is more than doubled.
The radius of the photosphere $R_{\rm f}$ cannot expand indefinitely because the
column density along the radial direction decreases with the expansion of the
gas giant. For self similar expansions, the total opacity of the gaseous
envelope is $\tau_{\rm tot} \propto R_{\rm exp}^{-2}$ where $R_{\rm exp}$ is an
expansion factor. For sufficiently large $R_{\rm exp}$, the envelope becomes
optically thin and the total energy can no longer be conserved. Rapid radiative
loss can reduce $W^\prime$ below zero. This transition is equivalent to that
from Sedov to snowplough phase of supernova shells. The envelope would be
totally disrupted if $W^\prime$ is sufficiently large that at the transition
phase, the outflow has already acquired a velocity in excess of the escape
speed. Otherwise, a fraction of the envelope may be retained in the form of
``fall back'' gas. 

Our models indicate that partial retention of a gas giant's envelope after a
giant embryo impact requires a narrow range of energy deposition. This energy
is generally unattainable from embryos and super-Earths near the sites of gas
giant formation. Thus the mass of long-period gas giants is generally retained
while their heavy elemental content is being enriched by GIM's.

\begin{table*}
\centering
\begin{tabular}{ccccccccc}
\hline\hline
 Model &   $M_{\rm I}$ &  $M_{\rm I,c}$ & $a$ & $v_{\rm imp}/v_{\rm esc,2}$ &  $M_{\rm f,c}$  & $M_{\rm f,g}$  & $R_{\rm f}/R_{\rm i}$  & $R_{\rm f}/R_{\rm S}$ \\
    &  ($M_\oplus$)  &  ($M_\oplus$) & (AU) &  & ($M_\oplus$) & ($M_\oplus$) & &  \\
\hline 
LA3   &   25  & 25 &  0.04  & 3.0  & 35  &  0   & -    & -      \\
 LA3a  &   25  & 25 &  0.1   & 3.0  & 35  &  0   & -    & -      \\
 LA3b  &   25  & 25 &  5.0   & 3.0  & 35  &  0   & -    & -      \\
 LB3   &   10  & 10 &  0.04  & 3.0  & 20  &  0   & -    & -      \\
 LB3a  &   10  & 10 &  0.1   & 3.0  & 20  &  90  & 2.38 & 2.61   \\
 LB3b  &   10  & 10 &  5.0   & 3.0  & 20  &  90  & 1.68 & 1.73 \\
 LC2   &  100  & 10 &  0.04  & 1.4  & 20  &  0   & -    & -      \\
 LC2a  &  100  & 10 &  0.1   & 1.4  & 20  &  0   & -    & -      \\
 LC2b  &  100  & 10 &  5.0   & 1.4  & 20  &  0   & -    & -      \\
 \hline
 \end{tabular}
 \caption{Comparisons of LHD models at different semi-major axes\label{tab:LHDCOMP}. 
The impacted model is a 100 $M_\oplus$ gas giant with a 10 $M_\oplus$
core. All the symbols are identical to the previous tables.}
\end{table*}

\subsection{De-synchronization of the envelope and further inflation}

If a state of spin-orbit synchronization has already been accomplished
by tidal evolution prior to the giant impact, both off-center
collisions and the expansion of the envelope would introduce
asynchronous spin and perhaps modest orbital eccentricity in general.
As an example, we simulate in model SC1b , the merger event of two
identical planets during which their cores also coalesce. Although
the envelope experienced a more significant expansion, most of the
envelope remains intact. In this case, the final spin rate would
differ substantially from its initial values.  This departure from
synchronous rotation may have observable consequences for atmospheric
flows \citep{Rauscher:2014pf} and thermal evolution.

Subsequent tidal dissipation would lead to an additional source of
heating for the planet \citep{Dobbs-Dixon:2004fu} such that

\begin{equation}
{\dot E}_{\rm t} =
{ G M_\ast M_{\rm p} e_{\rm p} ^2 \over a ( 1 - e_{\rm p} ^2) \tau_e}
+ {\alpha_{\rm p} M_{\rm p} R_{\rm p}^2 (n_{\rm p} - \Omega_{\rm p}) ^2 \over \tau_\Omega} \,,
\end{equation}
where $\tau_\Omega= (7 \alpha_{\rm p} / 2) (R_{\rm p}/a)^2 \tau_e$ is the
synchronization time scale and $\tau_e$ is the eccentricity damping time scale.
The rate of energy dissipation depends not
only on the differential angular frequency but also on the radius and
internal core-envelope structure of the planet through its $Q_{\rm p}
^\prime$ value \citep{Ogilvie:2004rw}. Although the amount of
extractable energy in the planet's spin is limited, for a brief
duration $\tau_\Omega$, the rate of tidal dissipation due to
synchronization is comparable to that due to circularization. Under
some circumstances, the planet's envelope would further expand and
perhaps overflow its Roche lobe, if the tidal dissipation rate exceeds
the energy loss rate at the planet's surface \citep{Gu:2003ez, Gu:2004km}.
We shall present an analysis of this possibility elsewhere.

\subsection{Consequence of giant impacts on HD\,149026}

Based on the results of the previous subsections, we propose the
progenitor of HD\,149026b formed with a Saturn-like internal
structure. When it migrated to the proximity of its host star, it may
have merged with another short-period gas giant or a population of
close-in super earths. If these collisions occurred with relatively low
impact speeds, phase transition of the core material would be avoided.
Nevertheless, heavy elements in the core would mix with an inflated
envelope. Subsequent tidal evolution provides additional heating which
may have led to the partial loss of its gaseous envelope through
Roche-lobe overflow. We note that HD\,149026b is less massive than most
known hot Jupiters and could have lost some mass. The metal-rich
debris material would stream into the host star and significantly
increase the metallicity of its shallow outer envelope
\citep{Li:2008vn}. Angular momentum transfer between the planet and
the debris stream would also enlarge the planet's orbital semi major
axis and Roche lobe so that the envelope loss would be limited. On a
time scale comparable to the age of its host star, this planet would
contract to its present-day radius.

We anticipate common occurrence of similar type of merger events. Since it is
difficult to assess the optimum initial conditions for this model, we now
present a series of simulations with a HD\,149026b-like planet as our initial
model. The total mass of the initial model is 110-$M_{\oplus}$, and the core
mass is around 73-$M_{\oplus}$. According to the observations of HD\,149026b
\citep{Sato:2005uq}, the initial model has $R_{\rm p}=0.73 R_{\rm J}$, $a=0.042$ AU, and
$T_{\rm e}=1500$ K accordingly.

We calculate models struck by embryos with different masses and speeds.
The model parameters and results are listed in Table~\ref{tab:LHDBIGCORE}.
For the models struck by a 10-$M_\oplus$ or a 25-$M_\oplus$ embryo with
low speed ($v_{\rm imp}/v_{\rm esc, 2}=1.0$), there is still inflation of its gaseous
envelope. For larger impact velocity $v_{\rm imp}/v_{\rm esc, 2}=1.4$, the gaseous
envelope of the model impacted by 25-$M_\oplus$ embryo has already expanded
beyond its $R_{\rm R}$.

When comparing the two series of models (Saturn-like and
HD\,149026b-like models), we find that Saturn-like models are more
resilient in retaining the gas envelope during the impacts. Even
though some of the Saturn-like models (LA2) are hit by impactor
with masses several times greater than that in the catastrophic
HD\,149026b-like model LB2S, there is no significant mass loss in these
models. This dichotomy may be explained by the following reasons.
In the calculations, we assume that all the impactors can reach the
core, with ablated mass and gravitational energy deposited in a region
around the core. Thus, the deposited energy by the impactor is not
only function of the mass, but also related its path and the internal
structure of the target. For
the same amount of material deposited at the same location, the
deposited energy would be larger in the HD\,149026b-like models than
that in the Saturn-like models, because the former have much larger
solid core than the latter. And a side effect of the massive
impactors is that the Hill radius will increase with the increasing
mass of impactor, which will make it more difficult for the envelope
to expand out of its Hill radius. Furthermore, the Saturn-like models
have very massive gaseous envelope compared with the HD\,149026b-like
models. They obviously require much larger energy deposition to
overcome the gravitational binding for the envelope to escape out of
the Hill radius.

\begin{table*}
\centering
\begin{tabular}{ccccccccc}
\hline\hline
 Model &  $M_{\rm I}$ &  $M_{\rm I,c}$  & $v_{\rm imp}/v_{\rm esc, 2}$ &   $M_{\rm f,c}$  & $M_{\rm f,g}$  &  $M_{\rm f,g}^{\rm N}$  & $R_{\rm f}/R_{\rm i}$  & $R_{\rm f}/R_{\rm S}$ \\
    &  ($M_\oplus$)  &  ($M_\oplus$) &  & ($M_\oplus$) & ($M_\oplus$) & ($M_\oplus$) & & \\
 \hline 
 LA1S  &   25  & 25 &  1.0  & 98  &  37 & 90  & 2.00  & 1.76   \\
 LA2S  &   25  & 25 &  1.4  & 98  &  0 & 90  & - & -   \\
 LA3S  &   25  & 25 &  3.0  & 98  &  0   &  0  & -    & -      \\
 LB1S  &   10  & 10 &  1.0  & 83  & 37  & 90  & 1.13 & 1.00   \\
 LB2S  &   10  & 10 &  1.4  & 83  &  37 & 90  & 1.47 & 1.29   \\
 LB3S  &   10  & 10 &  3.0  & 83  &  0  &  0  & -    & -  \\
 \hline
 \end{tabular}
 \caption{LHD models of HD\,149026b-like planet impacted by embryos
with different masses\label{tab:LHDBIGCORE}. 
The initial model has a total mass of 110
$M_\oplus$ and a 73 $M_\oplus$ core. In all of these models, the
semi major axis of the impacted planet is set to be 0.042 AU.
The results for collisions onto Saturn-like planet are also listed
as $M_{f,g}^N$ for comparison.}
\end{table*}

\section{Summary and Discussions}
\label{sec:summary}

\subsection{A brief summary}

During the formation and dynamic evolution of the planetary system,
there are many avenues of GIMs by embryos onto hot Jupiters.
The supply of these large solid building blocks includes:
1) planetesimals along the paths of migrating gas giants, 
2) super-Earths stalled at barriers outside the
asymptotic destiny of hot Jupiters, 
3) terrestrial bodies cleared by
sweeping secular resonances and dynamical instabilities, and 
4) short-period super Earths subjected to tidal orbital evolution. 

There are also several processes which can lead to the merger of gas
giants: 
1) orbit crossing triggered by run-away migration, 
2) congregation of hot Jupiters near their host stars, and 
3) long-term dynamical instability in the stellar proximity. 

Based on these consideration and the observation that a large fraction
of stars bear super-Earths and a large fraction of known gas giants
are members of multiple-planet systems, we infer common occurrence of
close encounters among these relatively massive gaseous and solid
planets. These events can lead to the observed Rayleigh distribution
of gas giants' orbital eccentricity \citep{Zhou:2007fk, Juric:2008uq,
Chatterjee:2008gd}.

\subsection{Potential outcomes of GIM events}

Here, we suggest that some of these close encounters results in
GIMs. The energy released during such collisions may lead to
expansion of the planets' envelope. If this GIM scenario is applicable
for the origin of the exceptional inflated close-in planets, their
fraction (relative to the normal-size planets) would correspond to the
duty cycle of a planet's expanded state following a major impacting
event. On the Kelvin-Holmheltz time scale ($\sim 10^8$ yr), the
inflated planets return to their initial state of thermal equilibrium. 

Under the assumption that the residual planetesimals and embryos are
sufficiently depleted that their characteristic collision time scale
with close-in gas giants is comparable to the age of their host stars,
the population of inflated planets is expected to be an order of
magnitude less than that of the normal size gas giants. This inference
is consistent with the relatively small fraction of inflated to normal
close-in gas giants.

In the proximity of their host stars, a substantial amount of the
envelope gas may be lost, after each giant impact, through Roche-lobe
overflow \citep{Gu:2004km}, leading to a large metallicity enhancement of
the planets. The intermediate-mass merger products may account for a 
population of planets found in the ``planetary desert'' which was predicted
by the population synthesis models \citep{Ida:2004ko, Howard:2010bs,
Ida:2013fv}.  The metal-rich planetary debris may also be accreted by
their host stars, leading to modest enrichment of the stellar
outermost envelope \citep{Li:2008vn}. These events of planetary
coalescence and the pollution of their host stars are unlikely to
occur if gas giants are formed through gravitational instability. The
verification of this GIM scenario may provide a distinguishing test
for these competing scenarios of planet formation.

For planet with longer orbital periods, tidal evolution toward spin synchronization is weak, angular momentum deposited by impacts can be reserved. In the future, oblateness measurements may be used to constrain impact history of gas giant planets.

\subsection{HD\,149026b}

In this paper, we considered the possibility that the large core of
hot Jupiter, HD\,149026b formed through giant impacts onto a gas giant
planet with either residual proto-planetary embryos or merger with
other gas giant planets in the proximity of its host star. In order
to examine the outcome of these collisions, we carried out a series of
numerical simulations with a SPH algorithm. These calculations
demonstrate gentle collisions always lead to coalescence whereas
nearly parabolic direct collisions can result in a preferential loss
of gaseous envelope material while the heavy elements in the core are
mostly retained. But, most of the core material may not merge or be
retained as a consequence of oblique high-velocity collisions.

We assume that a substantial fraction of the original residual
planetesimals in the neighborhood of HD\,149026b may have been scattered
into its host star which is an F star with a very shallow convective
layer. We carried out evolution calculations to take into account of
the effect of stellar pollution. We suggest that in this and other F
star with planets, the extent of stellar pollution by residual
planetesimals may be tested by an accurate determination of their
mass, luminosity, and effective temperature. Such a determination
will provide a firm support on the once existence of
terrestrial-planet-building material.

\section*{Acknowledgments}
We thank Drs F. Adams, E. Asphaug, P. Bodenheimer, J. Guillochon, 
T. Guillot, Y. Hori, S. Ida, M. Kouwenhoven and G. Laughlin for useful conversations. We also thank the referee David Stevenson for constructive comments. The software used in the hydrodynamic simulations was in part developed by the DOE-supported ASCI/Alliance Center for Astrophysical Thermonuclear Flashes at the University of Chicago. Computations were performed on the Laohu computer cluster at NAOC and the Hyades clusters at UCSC.
This work is supported by UC/Lab Fee grants. S.-F. L. was  sponsored by NASA NNX13AR66G "Collisional Accretion of Similar Sized Bodies". This work is also
supported by the Astronomy Unit's STFC Consolidated Grant.

\newpage
\bibliographystyle{mn2e}
\bibliography{ref}\label{sec:ref}
\end{document}